\documentclass[useAMS,usenatbib]{mn2e}  
\usepackage{epsfig}

\def\mc{M_{\rm c}}  
\def\rc{R_{\rm c}}

\def\tc{T_{\rm c}}  
\def\nw{n_{\rm w}} 
\def\vw{v_{\rm w}}  
  
\def\tw{T_{\rm w}}  
\def\lx{L_{\rm X}}   
\def\lo{L_{\rm O VI}}   

\def\spose#1{\hbox to 0pt{#1\hss}}     
\def\lta{\mathrel{\spose{\lower 3pt\hbox{$\mathchar"218$}}     
     \raise 2.0pt\hbox{$\mathchar"13C$}}}     
\def\gta{\mathrel{\spose{\lower 3pt\hbox{$\mathchar"218$}}     
     \raise 2.0pt\hbox{$\mathchar"13E$}}}     
  
\newcommand{\eg}{{\rm e.g.\ }}  
  
\newcommand{\ie}{{\rm i.e.\ }}  
\newcommand{\cm}{{\rm\thinspace cm}}  
\newcommand{\km}{{\rm\thinspace km}}  
\newcommand{\g}{{\rm\thinspace g}}  
\newcommand{\pcc}{\hbox{$\cm^{-3}\,$}}  
\newcommand{\gpcc}{\hbox{$\g\cm^{-3}\,$}}  
\newcommand{\s}{{\rm\thinspace s}}  
\newcommand{\yr}{{\rm\thinspace yr}}  
\newcommand{\erg}{{\rm\thinspace erg}}  
  
\newcommand{\pyr}{\hbox{\yr$^{-1}$}}  
\newcommand{\ergps}{\hbox{$\erg\s^{-1}\,$}}  
  
\newcommand{\kmps}{\hbox{$\km\s^{-1}\,$}}  
\newcommand{\pcmsq}{\hbox{$\cm^{-2}\,$} }

\newcommand{\K}{{\rm K}}

\newcommand{\Mdot}{\hbox{$\dot M$}}

\title[Conductive clouds in supernova-driven superwinds]{The dynamics and high-energy emission of conductive gas clouds in     
supernova-driven galactic superwinds}     
     
\author[Marcolini et al.]   
       {A.~Marcolini$^1$, D.K.~Strickland$^2$, A.~D'Ercole$^3$, T.M.~Heckman$^2$, and C.G.~Hoopes$^2$\\   
       $^1$ Dipartimento di Astronomia, Universit\`a di Bologna,  
       via Ranzani 1, 44127 Bologna, Italy. \\   
       $^2$ Department of Physics and Astronomy, Jhons Hopkins University,  
            3400 N.~Charles St., Baltimore, MD 21218, USA.\\  
       $^3$ Osservatorio Astronomico di Bologna,  
       via Ranzani 1, 44127 Bologna, Italy.}

\date{Accepted ..., Received ...; in original form...}    
    
\pagerange{\pageref{firstpage}--\pageref{lastpage}}    
\pubyear{2005}    
    
\begin{document}    
    
\maketitle    
    
\label{firstpage}    
    
\begin{abstract}   
  
Superwinds from starburst galaxies are multi-phase outflows  
that sweep up and incorporate ambient   
galactic disk and halo gas. The interaction of this  
denser material with the more diffuse hot  
wind gas is thought to give rise to the O{\sc vi} emission and absorption  
in the far Ultraviolet (FUV) and the  soft thermal X-ray emission observed in 
superwinds. In this paper we present high-resolution  
hydrodynamical models of warm   
ionized clouds embedded in a superwind, and compare the  
O{\sc vi} and soft X-ray properties to the existing   
observational data. These models include thermal conduction,  
which we show plays an important role in   
shaping both the dynamics and radiative properties of the resulting  
wind/cloud interaction. Heat conduction   
stabilizes the cloud by inhibiting the growth of Kelvin-Helmholtz  
and Rayleigh-Taylor instabilities, and   
also generates a shock wave at the cloud's surface that  
compresses the cloud. This dynamical behaviour   
influences the observable properties. We find that while O{\sc VI}  
emission and absorption always arises in cloud   
material at the periphery of the cloud, most of the soft X-ray arises  
in the region between the wind bow shock   
and the cloud surface, and probes either wind or cloud  
material depending on the strength of conduction and the relative   
abundances of the wind with respect to the cloud.  
In general only a small fraction ($\la 1$\%) of the wind   
mechanical energy intersecting a cloud is radiated away at UV and X-ray 
wavelengths, with more wind energy going into accelerating the cloud. 
Clouds in relatively slow cool winds radiate a larger 
fraction of their energy, which are  
inconsistent with observational constraints. 
Models with heat conduction at Spitzer-levels are found to produce  
observational properties 
closer to those observed in superwinds than models with no thermal 
conduction, in particular in 
terms of the O{\sc vi}-to-X-ray luminosity ratio, but 
cloud life times are uncomfortably short ($\la 1$ Myr) compared to 
the dynamical ages of real winds. We experimented with  
reducing the thermal conductivity for one set of model parameters, 
and found that even when we reduced conduction by a factor of 25  
that the simulations retained the beneficial hydrodynamical stability and 
low O{\sc vi}-to-X-ray luminosity ratio found in the Spitzer-level  
conductive models, while also having reduced evaporation rates. 
Although more work is required to simulate clouds for  
longer times and to investigate cloud acceleration and thermal  
conduction at sub-Spitzer levels in a wider range of models, we conclude 
that thermal conduction can no longer be ignored 
in superwinds.

\end{abstract}   
   
   
\section{Introduction}   
\label{sec:introduction}  
  
Superwinds are multi-phase, loosely-collimated   
galaxy-sized outflows with measured velocities in excess   
of several hundred to a thousand kilometers per second,  
driven from galaxies experiencing intense recent or ongoing  
star-formation, \ie starburst galaxies.  
These superwinds  
are common in the local Universe, occurring in nearly all  
galaxies classified as undergoing starburst activity  
\citep{lehnert95,heckman98},  
and appear ubiquitous among the Lyman break  
galaxies at redshift $\sim 3$ where they are blowing $\sim 100$ kpc  
sized holes in the inter-galactic medium   
\citep[\eg][]{pettini01,adelberger03}.  
  
Superwinds are believed to be driven by the thermal  
and ram pressure of an initially very hot ($T\sim 10^{8}$ K),  
high pressure ($P/k ~ \sim 10^{7} \K \pcc$) and  
low density wind, itself created from the merged  
remnants of very large numbers of core-collapse supernovae (SNe),  
and to a lesser  
extent the stellar winds from the massive stars, that occur  
over the $\sim 100$ Myr duration of a typical starburst event  
\citep{chevclegg}. The thermalized SN and stellar wind ejecta  
predicted by this model (which we shall call the wind fluid for  
convenience) is too hot and too tenuous to be easily observed (unless  
heavily mass-loaded, e.g. \citealt{suchkov96}),   
but hydrodynamical models of superwinds show that the wind fluid  
sweeps up and incorporates larger masses of ambient galactic  
disk and halo interstellar medium (ISM) into the superwind,  
material which is   
more easily detected observationally \citep{chevclegg,suchkov94,ss2000}.  
  
Indeed, the majority of nearby superwinds have been discovered  
using optical imaging and spectroscopy to identify outflow in the  
warm ionized gas (gas at $T\sim10^{4}$ K),  
in many cases directly imaging bipolar structures  
aligned with the host galaxy minor axis with kinematics indicative  
of outflow at velocities of $v_{\rm WIM} = $ a   
few $\times 100$ to $1000 \kmps$  
\citep*[\eg][]{axon78,mccarthy87,ham87,ham90,bland88,ham90,lehnert95,lehnert96a}. Superwinds from galaxies at high redshift  
are recognized by virtue of blue-shifted interstellar  
absorption lines from warm-neutral and warm-ionized gas  
species \citep[\eg][]{pettini00,fbb02,adelberger03},  
absorption features very similar to those seen  
in local starburst galaxies with superwinds  
\citep{phillips93,kunth98,gonzalezdelgado98,heckman2000}.  
Nearby superwinds have also been extensively studied in soft X-ray  
emission \citep[\eg][]{rps97,dwh98,strickland04a} which probes emission from  
hot gas with temperatures in the range $T \sim10^{6}$ to $10^{7} \K$,  
following their initial detection by  
{\it Einstein} X-ray observatory \citep*{watson84,fabbiano84}.  
Indeed, observations have demonstrated  
that all phases of the ISM found in normal late type galaxies are  
also incorporated into starburst-driven superwinds  
\citep[e.g., see the reviews of][]{blandhawthorn95,dahlem97}.  
  
Over the last 5 years new observations of superwinds, in particular  
spaced-based observations of O{\sc vi} absorption and   
emission from $T\sim10^{5.5} \K$ gas in the FUV,   
and thermal emission from $T \sim 10^{6-7} \K$ gas in X-ray regime,  
have substantially added to our understanding of the multi-wavelength  
properties of superwinds   
\citep{strickland00,heckman01,heckman02,strickland02,otte03,aloisi03,hoopes03}. 
These observations have shown that the majority of the  
soft X-ray emission in superwind is not due a volume-filling wind fluid,  
but that it and the O{\sc vi}-emitting and absorbing gas arise in   
some form of interaction between the wind fluid and denser  
ambient disk or halo ISM. This is in qualitative agreement with some of the  
analytical and theoretical models mentioned above. The observations,  
in particular those of O{\sc vi}, are not easily explained by  
the simple analytical models of wind-blown bubbles \citep{weaver77,maclow88}  
often applied to superwinds \citep[see \eg][]{heckman01}.  
 
\citet{chevclegg} first posited that both the optical  
nebular emission and the soft X-ray emission in superwinds  
\citep[as had been observed by][ for example]{axon78,watson84}  
is from shocks driven into clouds embedded in a high  
velocity wind of merged SN ejecta, primarily as  
the observed soft X-ray luminosities exceeded the predicted emission  
for the wind fluid. Later optical and low-spatial resolution   
X-ray observations  
were interpreted in the light of interactions between the  
wind fluid and ambient ISM, either with entrained cloud  
or with ambient disk or halo ISM at the walls of the outflow  
cavity \citep{mccarthy87,ham90,suchkov94}. In this picture clouds  
are either dense pre-existing structure over-run and incorporated  
into the wind, or generated at the cavity walls  
through the action of instabilities  
in the shell of swept-up ambient ISM (primarily either shell fragmentation   
through Rayleigh Taylor instabilities when the superbubble  
blows out of the disk, or dense gas ripped off the walls of the outflow  
cavity by Kelvin-Helmholtz instabilities).  
High resolution optical observations of superwinds show small  
scale clumps in the H$\alpha$ filaments of NGC 3079 and M82  
with sizes of $\sim 20$ -- $30$ pc, often associated with   
fainter elongated structures \citep{cecil01,ohyama02}. These are  
similar to the tadpole-like clouds found in simulations  
\citep{suchkov94,ss2000}.  
The basic wind/ISM interaction model for soft X-ray emission  
in superwinds is validated  
by modern high resolution X-ray observations \citep{strickland00,strickland02,strickland_vulcano,cecil02,schurch02,strickland04a}, which   
show the soft X-ray emission is structured very similarly to optical  
H$\alpha$ emission on scales from kpc down to as small as $10$ -- $20$ pc,   
with filaments, clumps and apparently limb-brightened walls.  
Much of this H$\alpha$ and soft X-ray structure is associated  
with large-scale filaments or the edges of the superwind,  
but examples of wind interactions with clouds  
or larger-scale obstacles exist, e.g. most-notably the M82 northern cloud   
\citep{devine99,lehnert99}.  
  
  
Returning to the issue of O{\sc vi} absorption and emission in superwinds,  
a wind/cloud interaction model is considered  
most plausible for explaining the absorption properties  
of neutral, warm photoionized and coronal phase gas in  
NGC 1705 \citep{heckman01}. The kinematics and column density of  
the O{\sc vi}-absorbing gas are inconsistent with the standard superbubble  
model, where the coronal phase gas arises in a conduction front  
at the walls of the superbubble   
shell \citep{castor75,weaver77}. In NGC 1705, the superbubble  
shell (traced by warm photoionized gas) has almost certainly ruptured,  
and is expanding more slowly than the O{\sc vi}-absorbing gas.  
The most likely explanation of the data is that the O{\sc vi}-absorbing  
gas is being generated as the hot gas from  
the interior of the   
bubble flows out past fragments of the ruptured superbubble shell.

In this paper we present high-resolution hydrodynamical models   
of warm ionized clouds embedded in a superwind, and compare  
the simulated O{\sc vi} and soft X-ray properties to the  
existing observational data.  
The interaction of a superwind with an embedded cool cloud is  
not the only form of wind/ISM interaction considered important  
in superwinds, but wind/cloud interactions undoubtedly  
do occur within superwinds and we shall focus on them in this  
paper.  
These models include the effect of thermal conduction,  
which we show plays an important role in shaping both the dynamics and  
radiative properties of the resulting wind/cloud interaction.  
Thermal conduction has long been suspected as being an important process  
in superwinds, but has not been included in hydrodynamical models  
of superwinds (with the exception of one model by \citealt{dercole99}).  
Simulations of cool clouds embedded in a hot fluid, including  
the effects of thermal conduction, have been performed before   
\citep{ferrara93,vieser00}.  
However, the models we present here are  
the first high-resolution simulations of an ISM cloud  
embedded in a hot, supersonic, high ram-pressure wind.  
  
The observational diagnostics available from any single waveband study of   
superwinds\footnote{For example, the diagnostic parameters  
that can be most robustly extracted from observation data are:  
Soft X-ray emission --- X-ray flux and surface brightness,   
emission-weighted mean temperature, and the apparent gas-phase abundance of O  
 (and some of the other alpha-elements) with respect to Fe. Kinematics  
and absolute element abundances, and associated parameters such as the  
emission integral can not currently be obtained or are too uncertain to  
make use of; O{\sc vi} emission --- fluxes, line-widths and simple  
kinematics; Optical emission --- fluxes and surface-brightnesses,   
spatially mapped kinematics and standard nebular diagnostics within  
the inner brighter regions of winds. The nebular emission appears  
dominated by stellar photo-ionization in the inner region of winds,  
but in many cases  
line ratios (in particular H$\alpha$/[N {\sc ii}]) become  
more ``shock-like'' at larger distances from the central starburst.}   
are too crude to strongly constrain among  
the many possible hypotheses available. However, multi-wavelength  
comparisons, \eg the spatial location of  
H$\alpha$ emission compared to the soft X-ray emission,  
have proved more useful and more robust   
\citep[\eg][]{mccarthy87,ham90,strickland00,strickland02}.  
The coronal phase gas responsible  
for O{\sc vi} emission and absorption in superwinds must either originate  
from the cooling of originally hotter gas, or is collisionally or   
conductively heated. In any case, O{\sc vi} is also a tracer of energetic  
collisional processes in superwinds in the same way X-ray emission is  
\citep{heckman02}.   
This is unlike the optical nebula emission  
in superwinds (\eg H$\alpha$ and [N{\sc ii}] emission),   
which can be photo-excited by massive stars or from  
X-ray irradiation, in addition to collisionally excited in shocks.  
The effect of extinction by interstellar  
material at the wavelengths of the O{\sc vi} doublet is equivalent  
to the extinction experienced by soft X-rays of energy $\sim0.3$ keV,  
somewhat mitigating the observational uncertainties in the  
absolute level of extinction.   
Oxygen is also the main coolant for plasma with temperature  
between $10^{5}$ and $10^{6}\K$, and second only in importance to  
Iron as a coolant in hotter gasses.   
The O{\sc vi}-absorbing or emitting  
gas is also the highest temperature  
phase in superwinds for which the velocity can currently be measured,  
yielding important clues to the relative kinematics of different  
phases in superwinds \citep[see \eg][]{heckman01}.  
Thus a comparison between the O{\sc vi} and soft X-ray properties  
of superwinds should hopefully prove as, if not more, instructive  
than the H$\alpha$/X-ray comparisons that have already been made.  
We have thus chosen to consider simultaneously O{\sc vi} and   
the soft X-ray observational properties of our modeled wind/cloud  
interactions, in  
order to better constrain which models, if any, provide a good match  
to the observed properties of superwinds.

In Section \ref{sec:numerical_method} we describe the numerical code  
employed to perform the models, the parameters of our chosen  
superwind and cloud model and how we calculate the O{\sc vi} and  
X-ray properties thereof. The results of the simulations   
performed are described in Section \ref{sec:results}. The  
implications of these results for understanding superwinds are  
discussed in Section \ref{sec:discussion}, followed by   
a summary of the results in Section \ref{sec:conclusions}.

\begin{table*}  
\begin{minipage}{130mm}   
\caption{Model parameters for the primary set of  
superwind/cloud interaction simulations.}  
\label{tab:cloud_in_wind_params}   
\begin{tabular} {lccrrrrrr}   
\hline   
Model & $ \rho_{\rm w}$  & $T_{\rm w}$ & $v_{\rm w}$  
	& $\chi$ & $\cal{M}$ & $T_{\rm BS}$ & $\sigma_{0}$   
	& $\tau_{\rm cc}$ \\   
       & (g cm$^{-3}$)   & (K)         & (km s$^{-1}$)   
	&        &           & (K)          &   
	& ($10^5$ yr) \\   
\hline   
T1LP(NC) & $1 \times 10^{-26} $ & $1 \times 10^{6}$ & 447    
	& 100 & 2.98 & $2.8 \times 10^{6}$ & 0.13  
	& 3.3 \\  
T1HP(NC) & $1 \times 10^{-26} $ & $1 \times 10^{6}$ & 1000   
	& 100 & 6.66 & $1.4 \times 10^{7}$ & 3.27  
	& 1.5 \\   
T5LP(NC) & $2 \times 10^{-27} $ & $5 \times 10^{6}$ & 1000   
	& 500 & 2.98 & $1.4 \times 10^{7}$ & 16.34  
	& 3.3 \\   
T5HP(NC) & $2 \times 10^{-27} $ & $5 \times 10^{6}$ & 2236   
	& 500 & 6.66 & $6.9 \times 10^{7}$ & 396.98  
	& 1.5 \\  
\hline  
\end{tabular}  
\par  
\medskip  
Model parameters are the same in models with and without (NC)   
heat conduction. See Section \ref{sec:numerical_method:model} for details.   
Parameters with subscript $w$ refer to the undisturbed  
wind fluid. The density contrast between the cloud and the undisturbed wind  
is $\chi=\rho_{\rm c}/\rho_{\rm w}$. The wind is super-or-hypersonic  
with respect to the cloud,  
with Mach number ${\cal M}= v_{\rm w}/c_{s, \rm w}$. The ratio  
of wind ram pressure to initial cloud or wind thermal pressure is  
$\sim 15$ in the low pressure (LP) models, and $\sim 75$   
in the high pressure (HP)  
models.  
The highest temperature  
in the wind bow shock around each cloud will be approximately  
the post-shock temperature for a strong shock, \ie   
$T_{\rm BS} \approx 3/16 \mu m_{\rm H} v_{\rm w}^{2}/k$.   
The dimensionless heat flux saturation parameter $\sigma_{0}$   
(equation~\ref{equ:sigma0})  
corresponding to this bow shock temperature   
(assuming the bow shock density  
$\rho_{\rm BS}=4\rho_{\rm w}$) is a relative measure of the   
\emph{initial} strength of  
the dynamical effects seen in the conductive models.  
The time scale for the cloud to be crushed by the shock driven   
into it is $\tau_{\rm cc}$ (equation~\ref{equ:tau_cc}).  
\end{minipage}   
\end{table*}

\section{The numerical method}   
\label{sec:numerical_method}  
 
As a first step in investigating wind/ISM interactions in superwinds,   
we explore the properties of simple model in which a single  
dense cloud (initially at rest) is embedded into a supersonic wind of hot   
gas. 
  
\subsection{Simulations}  
\label{sec:numerical_method:simulations} 
   
To run the simulations we use the 2D BOH (Bologna Hydrodynamics)   
hydro-code implemented with the thermal conduction. The code is based   
on a second-order upwind scheme \citep{bedogni86}, in which   
consistent advection \citep*{norman80} is implemented to   
reduce numerical diffusion.  We solve the usual hydrodynamic   
continuity equations.  We also include a further tracer variable   
representative of the cloud material which  
is passively advected; such a tracer allows us to compute the degree  
of mixing between the cloud and superwind material.  
  
To take into account the thermal conduction we adopt the operator  
splitting method. We isolate the heat diffusion term in the energy  
equation and solve the heat transport equation, alternatively along  
the $z$ and $R$ direction separately, through the Crank-Nicholson  
method which is unconditionally stable and second order accurate. The  
resulting system of implicit finite difference equations is solved  
according to the two-stage recursion procedure  
\citep[e.g.][]{ritchmyer67}. Following \citet{cowie77}, we  
adopt saturated fluxes to avoid unphysical heat transport in presence  
of steep temperature gradients (see below).

In all simulations we adopt 2D cylindrical coordinates. The central   
area of the computational domain is covered by an uniform grid with   
mesh size $\Delta R = \Delta z = 0.1$ pc. Beyond some distance to   
the center the linear mesh size increases geometrically in both   
directions with a size ratio of 1.07 between adjacent zones.   The   
grid edges are thus at large distances in order to avoid that    
possible spurious perturbations originating at the boundaries may   
affect the solution in the central region of interest.    
   
For the RE  models (see Section \ref{sec:numerical_method:model}) the total number of mesh points is 800 $\times$ 400  
($z \times R$), while the uniform grid region is covered by of $600 \times  
300$ points.  For the models with ram pressure the whole grid has $1200  
\times 400$ ($z \times R$) mesh points i.e. an extension of $\sim 3$ kpc   
in length by $\sim 1.3$ kpc in radius, while the uniform region has  
$1000 \times 300$ points.  
  
In all the models reflecting boundary conditions are enforced on the
$z$-axis.  In the RE models, outflow conditions are applied at all the
remaining boundaries. For the models with wind velocity $v_{\rm w}
\neq 0$, the superwind flows parallel to the $z-$axis and enters the
grid from right-side of the grid where inflow boundary conditions are
applied.
  
The simulations are run until the cloud, which is accelerated  
by its interaction with the wind, leaves the computational grid  
after $\sim 1$ to $1.5$ Myr.  
  
Radiative energy losses within the simulation are taken into account 
considering the cooling curve $\Lambda$ for Solar metallicities as 
parameterized by \citet{mathews78}. Models of stationary conductive 
clouds run assuming primordial abundances give essentially the same 
dynamical results. With less cooling, there is slightly more hot gas 
present ($\sim 10$\% more by mass), but essentially the soft X-ray and 
O{\sc vi} luminosity, and O{\sc vi} column densities are lower in 
proportion to the reduction in metallicity. Clearly in these 
simulations with strong conduction (\ie at Spitzer levels) the 
structure of conductive front is not controlled or significantly 
influenced by radiative cooling, as in that case we would have 
expected a very significant increase in the total mass of hot gas in 
the low abundance simulations. 
  
O{\sc vi} emission originates in a thin layer of the conductive front 
where the gas is at temperature $T \sim 3 \times 10^5$ K. It is 
crucial to have a numerical resolution good enough to capture this 
thin layer in the conductive front. With a cell size of 0.1 pc the 
O{\sc vi} emission from our simulations originates from a layer 6-7 
grid cells wide.  If we reduce the grid resolution by a factor of two 
(to 0.2 pc) cooling rates are higher, although the increase in $\lo$ 
and $\lx$ is a factor of $\leq 20$\% compared to 0.1 pc resolution 
simulations. This suggests that even at 0.2 pc resolution the 
conductive front is still marginally resolved at the O{\sc vi} layer, 
so that we are confident that the default 0.1 pc resolution 
simulations are numerical resolving the conductive interface. 
  
 
\begin{figure*}   
\begin{center}   
\psfig{figure=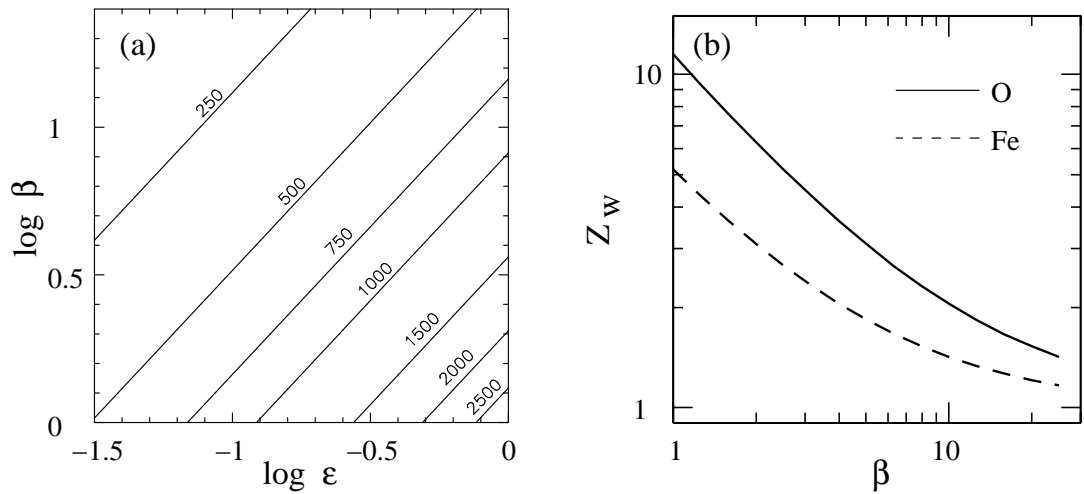}   
\end{center}   
\caption{(a) Superwind terminal velocity (in km/s)  
as a function of the efficiency of supernova energy thermalization 
$\epsilon$ and the factor by which the hot wind fluid is mass-loaded 
$\beta$.  (b) Oxygen and Iron abundance in units of standard Solar 
abundance as a function of $\beta$, assuming the ambient medium has 
Solar abundances.  See Section \ref{sec:numerical_method:model} for 
details.} 
\label{fig:mass_load}  
\end{figure*}   
  
\subsection{The wind/cloud model}  
\label{sec:numerical_method:model} 
  
 We make the following simplifying assumptions:  
\begin{enumerate}  
\item The clouds are initially spherical.   
\item The clouds are initially in pressure equilibrium with the ambient gas.  
\item The cloud self-gravity is neglected.   
\item Magnetic fields are neglected.
\item The gas is not allowed to cool below $T=10^4$ K.  
\item Non-equilibrium ionization effects on the cooling curve are neglected.   
\item The elemental abundances of both the wind and cloud material 
  are Solar. 
\end{enumerate}   

Assumptions (i) and (ii) are made to make simpler initial conditions.
However, they do not influence very much the results because, as it
will be described in Section 3, dynamical effects will quickly destroy
both the spherical shape and the pressure equilibrium of the cloud.
About assumption (iii), it is easy matter to show that, for the cloud
parameters adopted in this paper (see below) the cloud mass results
always lower than the cloud Jeans mass, even computed taking into
account external pressure \citep[see e.g.][]{shu92}. In spite
assumption (iv), we actually consider in Section 3.7 the possibility
of a reduced value of the thermal conductivity due to the presence of
magnetic fields. Point (v) assumes that the ionizing radiation field
due to the massive stars in the stellar burst is able to hold the
cloud ionized. Assumptions (vi) and (vii) are discussed in Section
\ref{sec:analysis} and Section \ref{sec:discussion:metal_abundances},
respectively.

The cloud properties, and the range in wind properties, we consider  
are consistent with observational  
and numerical studies of superwinds. At the location of the {\it FUSE}  
aperture used to survey O{\sc vi} emission in M82 and NGC 3079  
\citep[ Hoopes et al, in preparation]{hoopes03},   
the H$\alpha$-observations of \citet{ohyama02}  
and \citet{cecil01} show $\sim 20$ -- $30$ pc-scale clumps within the  
kpc-long filaments. The cloud-like structures visible in the  
highest resolution hydrodynamical simulations of \citet{ss2000}  
have diameters of $\sim 30$ pc, mass density $\sim 10^{-24} \gpcc$,  
and are embedded in wind fluid with a range of wind velocity, density  
and ram pressure similar to those we assume.  
  
We run different models of an evaporating cloud embedded into a galactic  
wind. In most of the models the cloud properties are the same: radius  
$\rc=15$ pc, temperature $\tc =10^4$ K and cloud mass density   
$\rho_{\rm c} = 10^{-24} \g \pcc$  
(proton number density $n_{\rm c}=0.42$ $\pcc$).  
The cloud  mass is $\sim 210 M_{\odot}$.    
We consider two sets of values for the density  
$\nw$ and the temperature $\tw$ of the superwind, both   
having the same thermal pressure $P=1.3 \times 10^{-12}$ dyne cm$^{-2}$:  
$(\rho_{\rm w},\tw)=(10^{-26} \,\g \pcc,10^6\,K)$,   
$(\rho_{\rm w},\tw)=(2 \times  
10^{-27}\, \g \pcc,5\times 10^6\, {\rm K})$. The cloud is thus  
initially in thermal pressure equilibrium with the wind in all the models.  
We further considered two  
possible values of the ram pressure exerted by the superwind on the cloud,  
namely $P_{\rm ram}=2 \times 10^{-11}$ dyne cm$^{-2}$ (low ram  
pressure models, model name suffix LP), and $P_{\rm ram}=10^{-10}$   
dyne cm$^{-2}$ (high ram pressure models, model name suffix HP).  
  
Each model is specified by the values of superwind temperature and ram  
pressure: for instance, the model T5LP is characterized by  
$(\nw,\tw)=(8.3\times 10^{-4}\,\pcc,5\times 10^6 \, {\rm K})$ and $P_{\rm  
ram}=2\times 10^{-11}$ dyne cm$^{-2}$ or, equivalently, a superwind velocity  
with respect to the cloud of  
$v_{\rm w}=1000$ km s$^{-1}$ (see Table~\ref{tab:cloud_in_wind_params}).    
To elucidate the role of the cloud size we also ran a model   
equivalent to model T1LP, but with a cloud radius  
$R_{\rm c} = 45$ pc and with a factor 3 lower resolution.  
  
For convenience we have chosen to treat the density, velocity and
metal abundance of the wind fluid as separate variables whose values
are not correlated.  However, theoretically these variables are not
independent of each other, as mass loaded winds will have lower
velocity and lower metallicity \citep[see e.g.][]{suchkov96}.  To
place the wind model parameters shown in
Table~\ref{tab:cloud_in_wind_params} in context we show the wind
terminal velocity and Oxygen and Iron abundances as a function of SN
energy thermalization efficiency $\epsilon$ and mass-loading $\beta$
in Fig.~\ref{fig:mass_load}.  The values shown were calculated using
the energy and mass return rates and elemental yields from SNe and
stellar winds from version 4 of Starburst99 \citep{leitherer99}.  We
assumed continuous star-formation of $1 M_{\odot}/ \yr$ at a time 30
Myr after the first star formation from gas of Solar abundances, but
the variation in $v_{\rm w}$ or $Z_{\rm w}$ from assuming a different
star formation history, time or initial metallicity is typically $\la
30$\%.  The mass injection rate in the starburst $\dot M = \beta
\times \dot M_{SN}$, where $\dot M_{SN}$ is the mass injection rate
due to SNe and stellar winds alone and $\beta$ is the degree of
mass-loading.  The terminal velocity in the hot wind fluid in
\citet{chevclegg} superwind model is $v_{\infty} = (2 \dot E / \dot
M)^{0.5}
\approx 2800 (\epsilon/\beta)^{0.5}$. To first order the density 
of the wind fluid scales as $\beta^{1.5}$. 
If the elemental abundance of any particular element $i$ 
is $Z_{i,SN}$ in the SN and stellar wind ejecta, and $Z_{i,ISM}$ in the 
ambient ISM, then metal abundance in the 
wind fluid will be $Z_{i,W} = \beta^{-1} \times (Z_{i,SN} +  
[\beta - 1] \times Z_{i,ISM})$. On the \citet{anders89} 
abundance scale $Z_{O,SN} \sim 11.5 Z_{O,\odot}$ 
and $Z_{Fe,SN} \sim 5.2 Z_{Fe,\odot}$. 
  
In order to disentangle the effect of the heat conduction from that
due to the ram pressure, we also ran models without heat conduction
(model suffix NC) as well as models of evaporating clouds in a static
hot medium (model prefix RE).  The model analogous to T5LP, but
without heat conduction, is called T5LPNC. This primary set of eight
models are summarized in Table~\ref{tab:cloud_in_wind_params}.  We
also ran two models equivalent to the model with conduction T5LP, but
with values of the coefficient of thermal conductivity that have been
reduced by factors 5 and 25 below the Spitzer value.  A further model
equivalent to model T1LP except with a cloud radius of $\rc = 45$ pc
was performed to investigate how the wind/cloud interaction depends on
cloud size.  The static medium models are characterized only by the
superwind temperature, thus, for instance, the model analogous to T5LP
is called RE05 (see Table~\ref{tab:models_at_rest}).

\begin{table}  
\centering  
\begin{minipage}{\columnwidth}  
\caption{Model parameters for models of a cold cloud in a stationary hot  
medium.}  
\label{tab:models_at_rest}  
\begin{tabular} {lccrr}  
\hline  
Model    & $\rho_{\rm w}$ & $T_{\rm w}$ & $\chi$ & $\sigma_0$    \\  
         &(g cm$^{-3}$)   & (K)         &         &              \\  
\hline  
RE01      & $1 \times 10^{-26} $  &  $1 \times  10^6$ & 100  & 0.07 \\  
RE05      & $2 \times 10^{-27} $  &  $5 \times  10^6$ & 500  & 8.59 \\  
RE10      & $1 \times 10^{-27} $  &  $1 \times  10^7$ & 1000 & 68.70 \\  
\hline  
\end{tabular}  
\par\medskip  
Parameters with subscript w refer to the static hot medium surrounding the  
cloud. The density contrast between the cloud and the hot medium is  
$\chi=\rho_{\rm c}/\rho_{\rm w}$. The dimensionless heat flux   
saturation parameter  
$\sigma_{0}$ is described in Section \ref{sec:results:static},   
equation~\ref{equ:sigma0}.  
\end{minipage}  
\end{table}  
  
\subsection{Analysis}  
\label{sec:analysis}
  
Calculations of the luminosity, 2-dimensional volume emissivities,  
and O{\sc vi} column densities from these models we performed   
separately from the simulations  
themselves. As stated above, we assume that the plasma is in  
collisional ionization equilibrium. The electron and ion temperatures  
in each computational cell are thus $T = (\mu \, m_{\rm H} \, P)/(k \,  
\rho)$, where $k$ is the Boltzmann constant, $P$ the thermal pressure,  
$\rho$ the total mass density, and $\mu m_{\rm H}$ is the mean mass
per particle. For the highly-ionized coronal and hot gas phases we are
interested in we use a value of $\mu m_{\rm H} = 1.02 \times 10^{-24}
\g \pcc$.  
  
In the low density limit the power emitted in some energy band $E$, in  
units of $\ergps$, from a elemental volume $\delta V$ is $\delta L_{E}  
= n_{e} \, n_{\rm H} \, \Lambda_{E}(T,Z) \, \delta V$, where  
$\Lambda_{E}(T,Z)$ is the emissivity at temperature $T$, for  
metal-abundance $Z$ in energy band $E$. For a given mass density  
$\rho$ and Solar abundances, then to good accuracy $n_{\rm H} = {\cal  
X}_{\rm H} \,  
\rho / m_{\rm H}$ and $n_{e} = 0.5 \, (1+{\cal X}_{\rm H}) \rho / m_{\rm H}$,  
where ${\cal X}_{\rm H}$ is the Hydrogen mass fraction. We use a value  
of ${\cal X}_{\rm H}=0.7057$. Making use of the assumed symmetry  
around the $z$-axis, we sum the luminosity over all cells within a  
cylindrical volume chosen to encompass the cloud and the majority of  
the wind bow-shock and any cloud fragments. We ignore any absorption  
of this emission from material on the computational grid.  
  
For the purposes of displaying which regions are the strongest O{\sc vi}  
and soft X-ray emitters we plot 2-dimensional maps of volume emissivity,  
i.e. emission per unit volume $\delta L_{E} / \delta V = n_{e} \,  
n_{\rm H} \, \Lambda_{E}(T,Z)$, in units of $\ergps \pcc$.  
  
The O{\sc vi} emissivities we use are based on the   
MEKAL hot plasma code \citep{mewe85,kaastra93,mewe95,liedahl95}.  
Note that the luminosity quoted is the sum of the two lines   
in the $\lambda = 1032$ \AA~ and 1038 \AA~doublet.   
  
The soft X-ray luminosities we quote are in the $E=0.3$ -- 2.0 keV energy  
band, chosen to correspond to the energy band used in the {\it Chandra} ACIS  
observations we compare to.  
The X-ray emissivities used are based on the 1993 update to  
the \citet{rs77} hot plasma code. Although the Raymond \& Smith code  
is no longer suitable for detailed X-ray spectroscopy, the broad-band  
emissivities it predicts are reasonably accurate.  
In the chosen energy band the  
difference between the broad-band soft X-ray emissivities given by  
the Raymond \& Smith code and the MEKAL code are $\sim 20$\% for  
gas with  
$10^{6} < T < 10^{7}$ K, and within a few percent for hotter gas.  
  
Given that the physical size of the observational {\it FUSE}  
$30\arcsec$ aperture depends on the distance to the observed galaxies  
(equivalent to $\sim 530$ pc for M82, and for $\sim 1.4$ kpc for NGC  
891, the closest and more distant galaxies we compare to),  
and that we do not know how many clouds actually lie within these  
regions, our quantitative comparison to the observational data will  
concentrate on the ratio of O{\sc vi} to soft X-ray emission, rather  
than the absolute emitted luminosities.  The physical volume  
represented by the full computational grid is much larger than the  
volume we would realistically expected to be occupied by only a single  
cloud in a superwind, and the X-ray emission from this volume is  
dominated by the wind, rather than wind/cloud interaction.  We  
therefore calculate the emission within a smaller cylindrical region  
judged large enough to encompass the cloud and the most luminous part  
of the bow shock. This region extends 20 pc upstream and 80 pc  
downstream of the initial center of the cloud, and out to a radius of  
50 pc.  
  
We also calculate the predicted O{\sc vi} column densities through the models,  
using an updated version of the code first used in \citet{heckman01}.  
Along any incremental path length $\Delta l$,   
the incremental O{\sc vi} column density   
$\Delta N_{\rm O VI} = \Delta l \, n_{\rm H} \, f_{\rm OVI}(T) \, A_{\rm O}$,  
where $n_{\rm H}$ is the local hydrogen number density and $A_{\rm O}$  
is the fractional abundance of oxygen atoms with respect to hydrogen  
($A_{\rm O}=8.51\times10^{-4}$ for Solar abundance in the \citealt{anders89}   
scale). In collisional ionization equilibrium, and at the low densities found  
in our simulations, the fraction of oxygen atoms in the O{\sc vi} state is  
purely a function of temperature $f_{\rm O VI}(T)$, the values for  
which we take from \citet{sutherland93}.  
  
The column density is evaluated along between 10 and 120 lines of  
sight through each  
simulation volume. For convenience we calculate lines of sight parallel  
to the $z$-axis (\ie parallel to the wind) and within a radius $R$,  
which may be larger or smaller than the actual radius of the cloud.  
The net O{\sc vi} column density quoted is the appropriately-weighted  
average O{\sc vi} column density over all lines of sight.   
Column densities calculated using larger radii are typically lower than the  
peak $N_{\rm O VI}$ column density that can be found using a single line of  
sight, or even the average over a radius $R \la R_{\rm c}$. Obviously the  
average O{\sc vi} column density will be reduced if many lines of sight  
do not intersect any gas in the temperature range in which the O{\sc vi}  
ion is abundant. Within the region of space probed by a typical   
{\it FUSE} LWRS aperture there are probably many clouds, but as their  
areal covering factor may in some cases   
be less than unity is worthwhile calculating  
$N_{\rm O VI}$ over radii $R \ga R_{\rm c}$.  
  
Cooling rates in all the hydrodynamical calculations assume Solar abundances   
(as given in \citealt{anders89}, which is the  
commonly used standard in X-ray astronomy). Although we track 
the contribution of cloud and wind material to the gas density in 
each computational cell and hence calculate the mean metal abundance 
within each cell, for the purpose of calculating the radiative 
cooling we assume $Z_{\rm w} = Z_{\rm c} = 1.0\times Z_{\odot}$. 
In reality  
cloud and wind material are likely to have different metal abundances, 
with the metal abundance of the wind depending on the degree of mass-loading. 
We shall demonstrate later that the rate of cooling within the wind is not 
dynamically significant. This allows us to post-facto consider  
the relative contribution of wind and cloud material to the X-ray 
and O{\sc vi} emission in cases when $Z_{\rm w} \not= Z_{\rm c}$. 
We shall discuss this issue further  
in Section \ref{sec:discussion}.

For each computational cell $i,j$ we assess the relative contribution 
by mass from wind material ${\cal R}_{ij} =  
\rho_{{\rm w}, ij}/( \rho_{{\rm w}, ij} + \rho_{{\rm c}, ij})$.  
We assume that the wind and cloud material 
in each cell mix to form a mixture with metallicity $Z_{ij} =  
(Z_{\rm w}\rho_{{\rm w}, ij}+Z_{\rm c}\rho_{{\rm c}, ij}) /  
(\rho_{{\rm w}, ij} + \rho_{{\rm c}, ij}) = 
Z_{\rm c} + {\cal R}_{ij}(Z_{\rm w} - Z_{\rm c})$. Which material  
dominates the emission from the system is then the luminosity-weighted 
sum of ${\cal R}_{ij}$,  
\begin{displaymath} 
<{\cal R}> = \frac{\sum L_{ij} {\cal R}_{ij}}{\sum L_{ij}}, 
\end{displaymath} 
taking into account the metallicity dependence of the luminosity due 
to each cell $L_{ij}$. The mean metal abundance that would be observed $<Z>$, 
for example in the X-ray band, would be the luminosity-weighted 
sum of $Z_{ij}$. Mathematically this is equivalent to  
$<Z> = Z_{\rm c} + <{\cal R}>(Z_{\rm w} - Z_{\rm c})$. 

Calculations following the exact ionization balance of the wind and
cloud material are currently beyond our capability.  We shall show
that radiative cooling is not dynamically important, but given our
interest in the FUV and X-ray absorption and emission predicted by
these models we shall briefly discuss the issue of non-equilibrium
effects.  Non-ionization equilibrium cooling calculations show that
the emissivity differs most from the equilibrium case for gas with a
temperature $T < 10^{6}$ K. Nevertheless, O{\sc vi} column densities
and Oxygen resonance line emissivities differ from the equilibrium
values by factors of at most $\sim 2$ -- 3
\citep{edgar86,sutherland93}. The latter paper demonstrates
that the O{\sc vi} ion is most abundant in non-equilibrium cooling
conditions at the same kinetic temperature as in collision
equilibrium.  Considering initially low-ionization gas being
conductively evaporated, the O{\sc vi} column density may be increased
by a factor of $\sim 3$ from the equilibrium case
\citep{weaver77,is04}, as the region of O{\sc vi} ions extends
into lower density and kinetically hotter material (the gas is 
under-ionized, in contrast to the over-ionized state in a cooling plasma). 
\citeauthor{weaver77} show that this leads 
to a minor reduction in soft X-ray luminosity, but it is not clear in
which direction and by how much the O{\sc vi} luminosity changes.
Thus we believe the cooling rates and column densities we calculate
assuming collisional equilibrium are most likely accurate to within a
factor $\sim 3$.  This magnitude of effect is not large enough to
alter our conclusions in a significant manner.

\section{Results}  
\label{sec:results}  
  
The main purpose of the present paper is the study of the  
O{\sc vi} and X-ray properties of conductively-evaporative cold clouds  
embedded in a hot superwind. However, before to present the results obtained  
for this scenario, it is instructive to consider  
the behaviour of the cloud in two  
extreme cases: $i$), one in which the evaporating cloud is at rest  
relative to the hot gas, and $ii$), one in which the cloud swept by  
the superwind does not suffer any evaporative mass loss (\ie thermal  
conduction is completely suppressed).  
  
  
\subsection{Evaporating clouds at rest in a static hot medium (RE models)}   
\label{sec:results:static}  
  
The first approach to the problem of the effects of thermal conduction   
on a cool cloud embedded in a hot medium has been undertaken by   
\citet{cowie77}.  
In the classical diffusion approximation of thermal conduction they   
obtained an analytical solution for the mass loss rate of an   
evaporating cloud given by   
\begin{equation}   
\dot{M}_{\rm cl}=4.34 \times 10^{-7} T_6^{5/2} {\rc}_{\rm ,pc} \;   
{\rm M}_{\odot} \, {\rm yr}^{-1},   
\label{equ:mdotcl}  
\end{equation}   
\noindent   
where $T_6=T/(10^6\, {\rm K})$ is the temperature of the ambient gas,   
and ${\rc}_{\rm ,pc}=\rc/(1\,{\rm pc})$ is the cloud radius.  In  
addition, they pointed out that the classical diffusion approximation  
breaks down when the mean free path of electrons becomes larger than  
the temperature scale (saturation). The relevance of this effect is  
quantified by the dimensionless parameter $\sigma_0$ defined as  
\begin{equation}   
\sigma_0=4.22\times 10^{-3}{T_6^2 \over n_{\rm H} {\rc}_{\rm ,pc}},   
\label{equ:sigma0}  
\end{equation}  
\noindent  
where $n_{\rm H}$ is the number   
density of hydrogen atoms in the ambient gas (not the number  
density within the cloud). In the   
saturated regime ($\sigma_0>1$) the evaporation rate is given by   
\citet{cowie81}:  
\begin{equation}  
\dot M=1.36 \dot M_{\rm cl} \sigma_0^{-5/8}.  
\label{equ:mdot_saturated}   
\end{equation}   
\noindent  
The above results are obtained in the assumption of steady, isobaric   
(and thus low Mach number) flow of the evaporating gas, and possible   
hydrodynamic effects that may arise in the system are neglected.   
\citet{mckee75} have analyzed the evaporation process in terms   
of a conduction front advancing into the cooler gas. They outlined an   
interesting analogy with the ionization fronts, and, just as in that   
case, the velocity of the front $v_{\rm cond}$ must be less than   
$0.5c_{\rm c}^2/c_{\rm w}$ (D-type front) or greater than $2c_{\rm w}$   
(R-type front), where $c_{\rm w}$ is the sound speed of the hot   
gas, and $c_{\rm c}$ is the sound speed of the cloud. In the case of an    
evaporating cloud, \citet{cowie77} have   
shown that $v_{\rm cond}\sim 2\sigma_0 c_{\rm c}^2/c_{\rm w}$ for   
classical evaporation.  Thus a critical value $\sigma_{\rm 0,cr}=0.25$   
exists beyond which the conduction fronts are D-critical, and   
dynamical effects are important.  Saturated evaporation fronts,   
instead, have $v_{\rm cond}\sim 1.12\sigma_0^{1/8} c_{\rm c}^2/c_{\rm   
w}$ and are always D-critical. In this case the hot pressurized gas   
behind the front expands, driving a shock wave which propagates through   
the cold gas ahead of the conduction front itself.   
  
The above conclusions apply under the assumption that radiative   
losses are negligible.   
If such losses exceed the conductive heat input, material   
condenses onto the cloud instead of evaporating. However, \citet{mckee77}   
have shown that this occurs for $\sigma_0<0.03$, a circumstance never met   
in our models.  
   
For our models RE01, RE05 and RE10 we have $\sigma_0= 0.07$, $\sigma_0= 8.59$   
and $\sigma_0= 68.7$, respectively. Hydrodynamic effects are thus   
expected to be absent in model RE01, but important in models RE05 and   
RE10. Actually, our simulations confirm this expectation.   
In model RE01 the cloud evaporates subsonically at the classical rate,   
without experiencing any compression. In models RE05 and RE10 shock waves   
develop and compress the evaporating cloud.

As we show in Section \ref{sec:results:evap_clouds},   
all the models of evaporating clouds  
dragged by the superwind develop shock waves driven by the conduction   
front; we thus describe here in detail the model RE05, in order to   
emphasize the cloud dynamics generated by the thermal evaporation. As   
a result of the energy transfer by thermal conduction, the pressure   
at the cloud surface rises and drives two opposite flows.  The   
evaporating gas moves outward subsonically while the inward flow is   
supersonic (Mach number ${\cal M} \ga 3$) and drives a shock   
wave. This different behavior is tied to the large temperature   
difference between the cloud and the hot gas, and the resulting   
difference in the sound speeds: $c_{\rm w}/c_{\rm c}$=22.4.    
  
\begin{figure}   
\begin{center}   
\psfig{figure=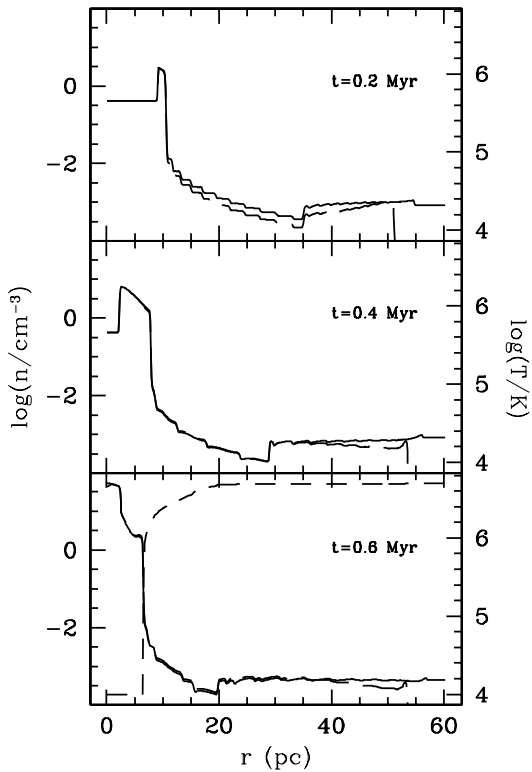}
\end{center}   
\caption{Density (solid line) and temperature (short dashed line) radial  
profiles as a function of time in  
the model RE05. The cloud center is located at $R=0$, while the cloud  
edge was initially at $R=15$ pc. The upper two panels show the  
evolution of the spherical shock induced by the thermal conduction  
while it is converging toward the center. The bottom panel displays  
the cloud structure after the shock has bounced back. The long dashed  
line represents the cloud gas density.}  
\label{fig:evolution}  
\end{figure}

The spherical shock originating at the surface compresses the cloud as 
it converges toward the center.  The evolution of model RE05 is 
illustrated in Fig.~\ref{fig:evolution}, where the cloud center is 
located at $r=0$.  The first two panels of Fig.~\ref{fig:evolution} 
show the cloud evolution at $t=0.2$ Myr and $t=0.4$ Myr, when the 
converging shock has not yet reached the center.  Given the strong 
radiative losses, the shock is isothermal.  The cloud temperature 
remains constant because it is not allowed to cool below its initial 
value of 10$^4$ K, and thus the cloud pressure varies in pace with the 
cloud density.  It is apparent that the post-shock density (and 
pressure) increases with time, as a consequence of the energy 
''focusing'' as the wave approaches the center. This effect is 
expected for an converging adiabatic shock 
\citep[\eg][]{landau87}, but remains valid also for a radiative  
shock, at least in the parameters range of our models. 
  
It is also evident that the density at the outer edge of the  
compressed shell, where the conduction front is located, decreases  
despite the overall compression of the cloud. In fact, given the  
negative value of the radial pressure gradient, with time the tail of  
the density profile lags more and more behind; the outer layers of the  
cloud progressively retard their collapse, and possibly revert their  
motion into an expansion. The density decrease at the conduction  
front strongly influences the evolution of OVI and X-ray cloud  
luminosities, as discussed in Section \ref{sec:results:static:xray}.

\begin{figure}   
\begin{center}   
\psfig{figure=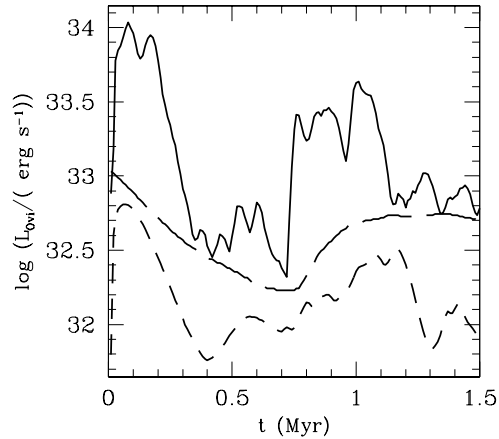}
\end{center}  
\caption{Time evolution of $\lo$ (solid line) and $\lx$ (dashed line) for 
the RE05 model. The cloud surface area $\rc^2$ is also shown in arbitrary  
units (long dashed line line).}  
\label{fig:lo_lx_re05}  
\end{figure}

At $t=0.43$ Myr the shock reaches the center and bounces back,  
re-expanding through the gas still moving toward the center. At this  
time the central density reaches $n\sim 2.1\times 10^3$ $\pcc$,  
nearly 5000 times larger than the initial cloud density.  
  
The lowest panel in Fig.~\ref{fig:evolution}  
shows the density and temperature  
profiles at 0.6 Myr. The density jump appearing at $r\sim 2.5$ pc   
indicates the position of the reflected shock. At this stage the   
central density of the cloud is a factor of 100 larger   
than the initial value.  
The cloud radius attains its minimum value $\rc = 5$ pc at $t\sim 0.7$  
Myr, only shortly before the reflected shock reaches the edge of the  
cloud at $t\sim 0.75$ Myr. This latter occurrence influences greatly  
the cloud emission as discussed in the next subsection.  
  
We point out that, as $\rc$ re-expands, the thermal energy flux  
through the cloud surface increases as well. At $t=0.95$ Myr, when the  
radius reaches the value of $\rc =12$ pc, conditions are restored for  
a new collapse driven by the heat conduction. This oscillatory  
behaviour is not found by \citet{ferrara93} who performed  
similar 1D simulations; in their models the cloud attains a steady  
configuration after the initial compression. The difference is due to  
the fact that in their models the cloud is allowed to cool to arbitrarily low  
temperatures, whereas we impose a minimum temperature  
to approximate photo-ionization heating from massive stars.  
  
\subsubsection{O{\sc vi} and X-ray emission in RE models}  
\label{sec:results:static:xray}   
  
As a gas element evaporates from the cloud, it quickly goes through a  
large interval of temperatures, ranging from $T \sim 3 \times 10^5$ K  
where the O{\sc vi} emission peaks, up to X-ray temperatures  
(several $10^6$ K). Thus significant O{\sc vi} and X-ray emission  
originates close to the cloud surface, and both are  spatially  
connected. For this reason the time evolution of $\lo$ and $\lx$ are  
coupled to the cloud dynamics, in particular to the variations of  
the cloud density at $\rc$ and to the evolution of the cloud size.  
  
\begin{figure}   
\begin{center}   
\psfig{figure=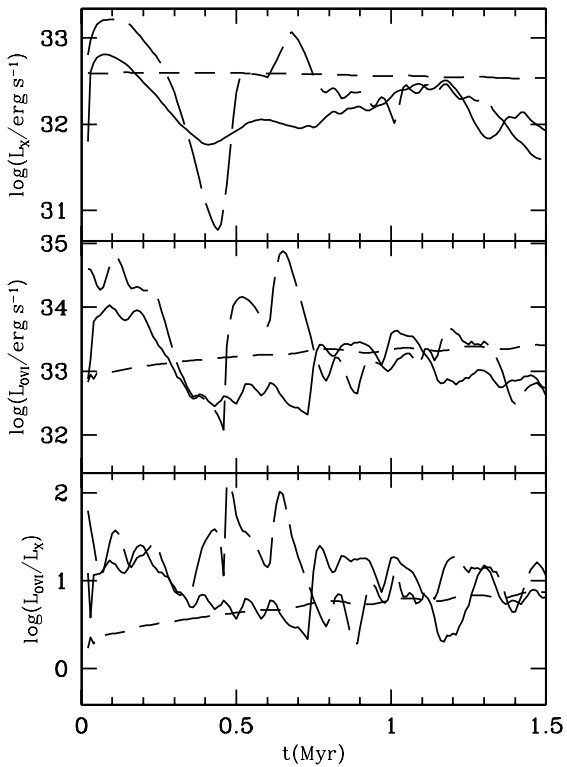}
\end{center}   
\caption{Time evolution of $\lx$ (upper panel), $\lo$ (middle panel) and   
log($\lo /\lx$) (lower panel) for all the rest models RE. Dashed line: RE01   
model; solid line RE05 model; long dashed line: RE10 model.}   
\label{fig:lo_lx_remodels}  
\end{figure}

Both $\lo$ and $\lx$ have similar temporal profiles, both coarsely  
modulated by the behaviour of $\rc$. This is illustrated by   
Fig.~\ref{fig:lo_lx_re05}, which shows the evolution of $\lo$, $\lx$ and  
$\rc^2$ for the model RE05. The sudden initial rise of  
$\lo$ and $\lx$ is due to the formation of the emitting layer at the  
cloud surface. Successively, the luminosity decreases in pace with the  
cloud radius; we point out, however, that the drop of $\lo$ and $\lx$  
can not be explained only on geometrical grounds as an effect of the  
reduction of the emitting surface area. Fig.~\ref{fig:lo_lx_re05} clearly  
illustrates that the luminosity drop is larger than that of  
$\rc^2$. Much of the luminosity reduction is due to a reduction of the  
gas density at the conduction front, a consequence of the dynamics  
of the imploding cloud, as previously discussed.

The sudden rise in both X-ray and O{\sc vi} luminosity   
occurring at $t=0.75$ Myr shown in Fig.~\ref{fig:lo_lx_re05}   
coincides with the arrival of the outwardly-propagating reflected shock   
at $R_{\rm c}$.   
This shock compresses the gas of the emitting layer located  
at $R_{\rm c}$, leading to a rapid increase in emission.  
  
In general, the evolution of $\lo$ and $\lx$ is affected by a  
considerable ``noise'' due to the energy continuously injected into the  
cloud by the conduction front.  Even if the conditions at the cloud  
edge are such that a strong shock can not be driven by the heat flux,  
density fluctuations are still generated which influence the  
luminosities.

\begin{figure*}   
\begin{center}   
\psfig{figure=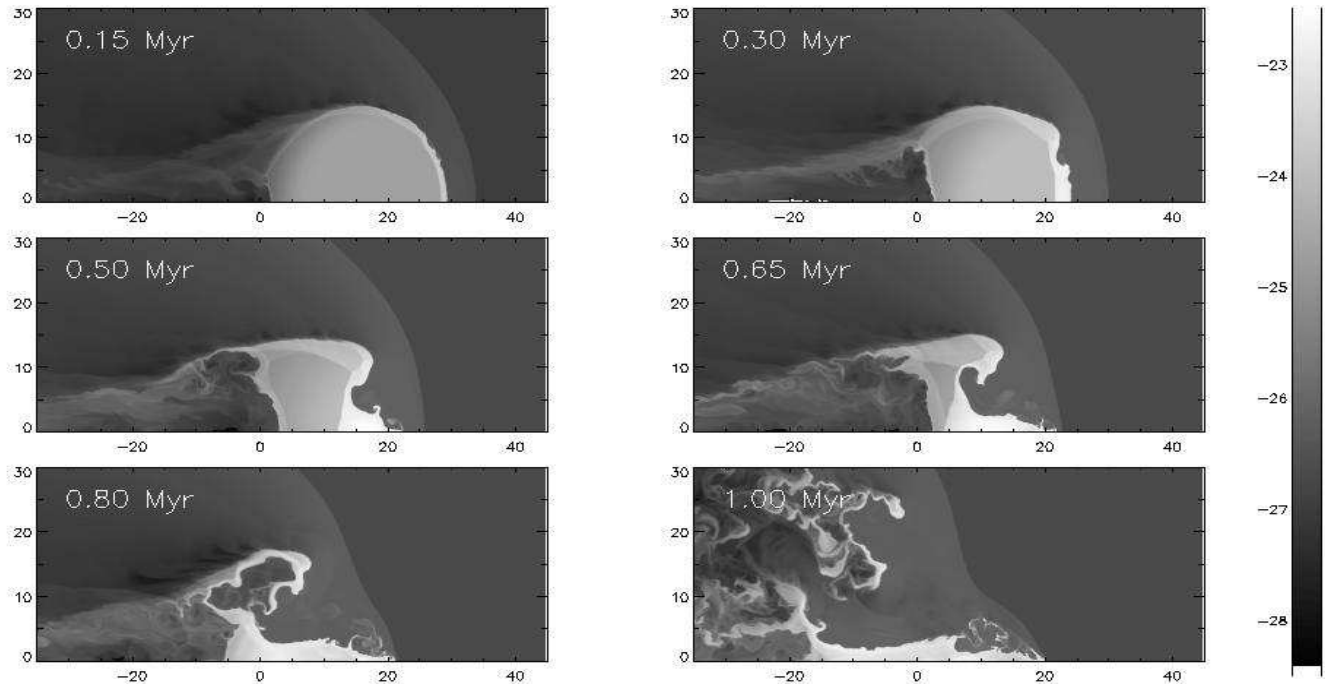}   
\end{center}  
\caption{Density distribution at different times of the cold cloud   
interacting with the hot tenuous superwind for the reference model   
T5LPNC (no heat conduction). The logarithm of  
the mass density (in units of $\gpcc$) is shown.  
The distances are given   
in pc. At the beginning of the simulation the cloud center is at $z=20$ pc.}  
\label{fig:ral55}  
\end{figure*} 

The luminosities $\lo$ and $\lx$ of the three models RE01, RE05 and RE10  
are compared in Fig.~\ref{fig:lo_lx_remodels}.  
In the model RE01 the dynamical effects are  
negligible, and the cloud suffers a steady evaporation. The rate of  
such evaporation is rather low, $\dot{M} \sim 6.4\times 10^{-6}$  
M$_{\odot}$ yr$^{-1}$, in good   
agreement to the analytical result presented in  
equation~\ref{equ:mdotcl}, and the cloud  
radius remains constant.  The O{\sc vi} flux increases slowly  
with time as the conduction front becomes thicker. The  
X-ray flux is not interesting for this model because the temperature  
range through the conduction front is obviously below $T_w=10^6$, and  
thus no significant X-ray emission occurs at the cloud edge in this  
case.  The constant level of $\lx$ shown in Fig.~\ref{fig:lo_lx_remodels}   
is essentially due to the surrounding hot medium.  
  
Compared to model RE05, the oscillations of $\lo$ and $\lx$ are  
accentuated in model RE10, where the heat flux is larger because of the  
larger temperature gradient at the cloud edge.  
  
In the bottom panel of Fig.~\ref{fig:lo_lx_remodels} the ratio  
$\lo/\lx$ is plotted. We will discuss how this ratio  
relates to the values observed in superwinds in Section \ref{sec:discussion}.

\subsection{Non evaporating cloud dragged by the superwind (NC models)}  
  
The interaction of a strong shock with a single cloud has been the
subject of many numerical studies. In the case of non-radiative clouds
a thorough analysis of this problem and a review of relevant literature
is provided by \citet{klein94}. Numerical simulations in the case of
radiative clouds have been presented by \citet{mellema02} and
\citet{fragile04}. In all these papers, after the passage of the shock the
cloud is embedded in a low-density wind, analogous to the situation
considered here.  However the model parameters in the papers just
mentioned are not appropriate for clouds embedded in superwinds.  We
thus run models of non-evaporating clouds (\ie without thermal
conduction) embedded in a superwind.
  
The purpose of these models is twofold: on one hand they represent an
useful check when compared to the results by other authors and, on the
other hand, they help to understand, by comparison with analogous
models with heat transfer, the role played by the conduction in the
general dynamics of the cloud.
  
In all our models (see Table~\ref{tab:cloud_in_wind_params})   
the superwind is super-or-hyper-sonic (Mach number ${\cal M}=3.2$ for models  
with low ram pressure and ${\cal M}=7.1$ for models with high ram  
pressure), despite the high temperature assumed for the volume-filling  
wind material.    
A bow shock forms around the cloud, while a shock is  
driven into the cloud with velocity $v_{\rm s}=v_{\rm w}\chi ^{-1/2}$,  
where $\chi=\rho_{\rm c}/\rho_{\rm w}$.   
The characteristic time $\tau_{\rm cc}$ for the cloud to be  
crushed by this shock is $\rc /v_{\rm s}$, \ie   
\begin{equation}  
\tau_{\rm cc}=\chi  
^{1/2} \rc/v_{\rm w}.  
\label{equ:tau_cc}   
\end{equation}    
This is the basic timescale governing the  
evolution of a cloud over-run by a blast wave or a wind,  
with cloud destruction expected after several  
crushing times \citep{klein94}. The cloud is fragmented by the action of both  
Kelvin-Helmholtz (K-H) and Rayleigh-Taylor (R-T) instabilities. The  
Richtmyer-Meshkov instability also contributes to the fragmentation,  
but it is less important because it grows linearly rather than  
exponentially.  
  
The K-H instabilities are due to the relative motion between the cloud   
and the hot, tenuous background gas.  The modes destroying the cloud   
are those whose wavelength is comparable to $\rc$, and their growth   
time is $\tau_{\rm KH}\sim \tau_{\rm cc}$ \citep[\eg][]{klein94}.    
This timescale estimate is a lower limit as with   
the formation of the bow shock around the cloud   
the gas actually flowing around the cloud edge has a   
velocity lower than that of the unperturbed superwind.   
  
The growth time of the R-T instability is $\tau_{\rm  
RT} \sim (\rc /a)^{1/2}$, where $a \sim \pi \rho_{\rm w} {\rc}^2 v^2_{\rm w}  
/ \mc$ is the cloud acceleration due to the superwind ram pressure. Again,   
the growth time is comparable to the crushing time, $\tau_{\rm RT}\sim   
\tau_{\rm cc}$. Again the quoted timescale is a lower limit.   
In reality the wind attempts to   
form a smooth flow around the cloud, and thus the acceleration  
$a$ is lower than the above estimate.    
  
The timescales given above hold in the case of an adiabatic evolution  
of the gas, however they are still adequate to roughly characterize  
the flow in the radiative case \citep{fragile04}. In fact,  
our radiative simulations substantially confirm the picture outlined above.   
  
In Fig.~\ref{fig:ral55} we show the density  
evolution of the low ram pressure  
model T5LPNC, for which $\tau_{\rm cc}=3.3 \times  
10^5$ yr.  The perturbations grow first close to the axis of symmetry  
of the cloud, where the R-T instabilities are more effective. Of  
course K-H instabilities are also present, and the actual growth time  
is $\tau=(\tau^{-2}_{\rm RT}+\tau^{-2}_{\rm KH})^{1/2}\sim \tau_{\rm  
cc}/\sqrt {2}$ \citep{lamb45}. The K-H instabilities are  
expected to develop faster near the tangential point, where the  
relative velocity is larger. However, an inspection to  
Fig.~\ref{fig:ral55} shows that this region is rather smooth; in fact  
the K-H modes originate at the cloud edge but grow downstream, where  
they are advected by the flow. For $t<t_{\rm cc}$, while the  
transmitted shock is still crossing the cloud, the flow of the  
superwind results in a low-pressure region at the rear of the cloud,  
into which the cloud material moves generating a rarefaction wave  
which propagates into the cloud. This low-pressure region produces  
also a sort of back-flow of the superwind gas, giving rise to a vortex  
ring responsible of the K-H instabilities which severely distort the  
back side of the cloud.  
  
Although part of the cloud is highly fragmented after a few crushing 
times, its core remains rather compact although highly distorted. At 
$t=1$ Myr the mass of the core is 47 $M_{\odot}$ (the original cloud 
mass was $\sim210 M_{\odot}$), while the rest of the cloud, as a 
consequence of the efficient radiative cooling, is leaving the grid 
downstream in form of dense filaments. 
  
Our results are intermediate between those expected for a 
non-radiative cloud and those obtained for a radiative cloud. In the 
former case the cloud would either be compressed, or be completely 
destroyed and diffused into the ambient medium (Klein et 
al. 1994). The radiative cloud, instead, breaks up into numerous, 
dense, cold fragments which survive for many dynamical timescales 
\citep{mellema02,fragile04}. In our simulations the 
fragmentation process is present but less efficient. This is due to
the fact that in our models the cloud is not allowed to cool below its
initial temperature of $10^4$ K, while in the papers quoted above the
minimum allowed temperature is $T=10$ K. We run a further simulation
(not shown here), analogous to the reference model T5LPNC, in which
the cloud is now allowed to cool down to 10 K. In this case radiative
losses are more effective and the fragmentation proceeds more
efficiently. The evolution is now very similar to that found in
previous works of \citet{mellema02} and \citet{fragile04}, and the cloud
is fragmentated in several small clumps after few dynamical timescale.
 
\begin{figure*}   
\begin{center}   
\psfig{figure=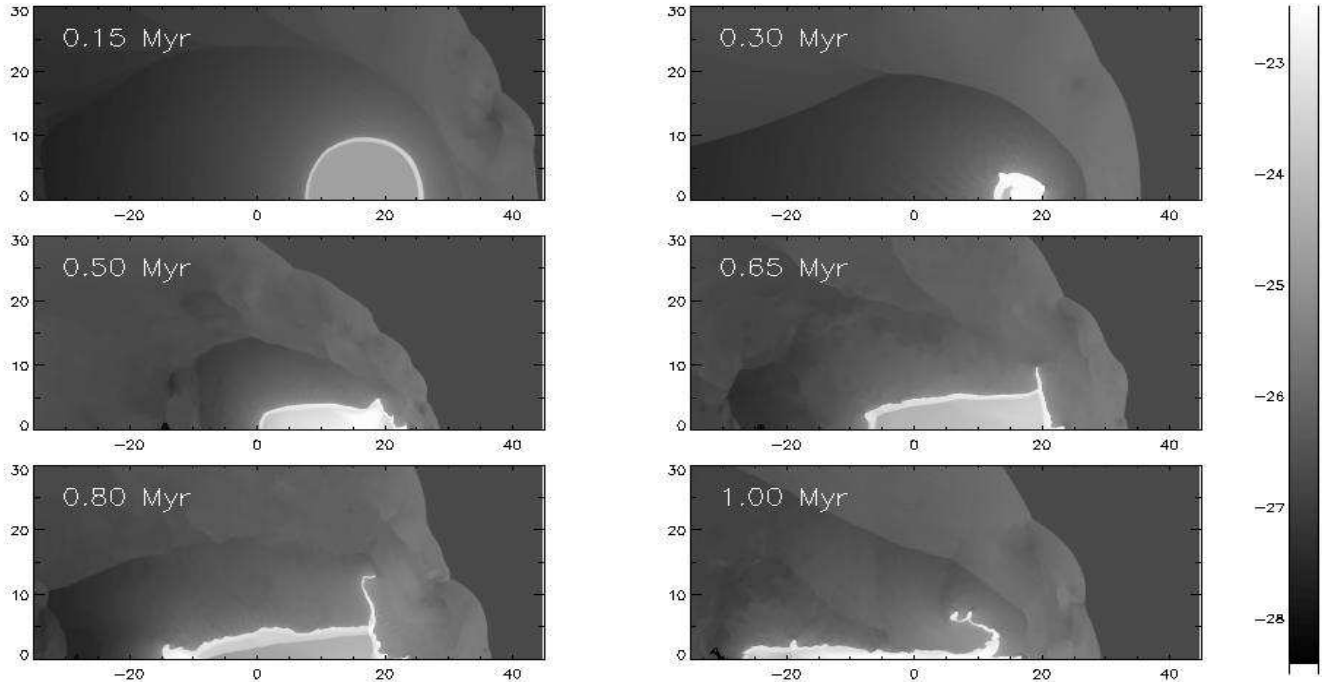}   
\end{center}   
\caption{Density distribution at different times of the cold cloud   
interacting with the hot tenuous superwind for the reference model  
T5LP including the effects of heat conduction.  The logarithm of the  
mass density (in units of $\gpcc$) is shown.  The distances are given  
in pc.  At the beginning of the simulation the cloud center is at  
$z=20$ pc.}  
\label{fig:rar55}  
\end{figure*}

\subsection{Evaporating clouds dragged by the superwind}  
\label{sec:results:evap_clouds}  
  
As already discussed, we found that in the RE models  
the dynamical effects linked to the thermal conduction become  
important when $\sigma_{0}>\sigma_{\rm cr}$. The post-shock gas facing  
the front of the cloud has a temperature $T_{\rm bs} \sim 1.38 \times  
10^7 (v_{\rm w}/1000 \kmps)^2$. As this  
temperature is always larger than the temperature $T_{\rm  
w}$ of the unperturbed superwind we assume, the effective initial   
value of $\sigma_{0}$ is larger than in the static case   
(see Table~\ref{tab:cloud_in_wind_params}). In  
particular, $\sigma_{0}$ is increased above $\sigma_{\rm cr}$ also in the  
models T1LP and T1HP (which otherwise would have $\sigma_{0}<\sigma_{\rm  
cr}$), and dynamical effects due to the thermal conduction are always   
present in all our models of clouds within a wind.  
  
In Fig.~\ref{fig:rar55} we show the evolution of the density for the   
reference model T5LP.   
This model develops a bow shock temperature $T_{\rm bs}  
\sim 1.3 \times 10^7$ K.  As expected, a converging shock due to the  
action of the thermal conduction forms all around the cloud (see the  
shot at $t=0.15$ Myr in Fig.~\ref{fig:rar55}), and the cloud radius  
has shrunk by a factor of three by  $t=0.30$ Myr.   
However, the shock formed at  
the leading edge of the cloud is stronger and moves faster through the  
cloud than the shock portion at the rear of the cloud. The reason for  
this is twofold: $i)$ the temperature in front of the cloud is larger  
than at its backside, and $ii)$ the superwind ram pressure  
contribution is also present at the front side.  The cloud continues  
to collapse till $\sim 0.4$ Myr, when the central core reaches a  
pressure high enough to stop the shrinking.  At this time the cloud  
re-expands.  The expansion occurs preferentially downstream, where it  
is not contrasted by the ram pressure of the incoming  
superwind.  With time the cloud looses completely its initial shape  
and assumes an elongated aspect. A successive compression starts at  
$\sim 0.75$ Myr. However, given the complicated cloud morphology and  
the complexity of the flow, the cloud compression is not uniform, but  
different regions shrink at different rates (see panel of  
Fig.~\ref{fig:rar55} at $t=0.8$ Myr). After 1 Myr the cloud assumes a  
filamentary shape along the symmetry axis.

\begin{figure*}   
\begin{center}   
\psfig{figure=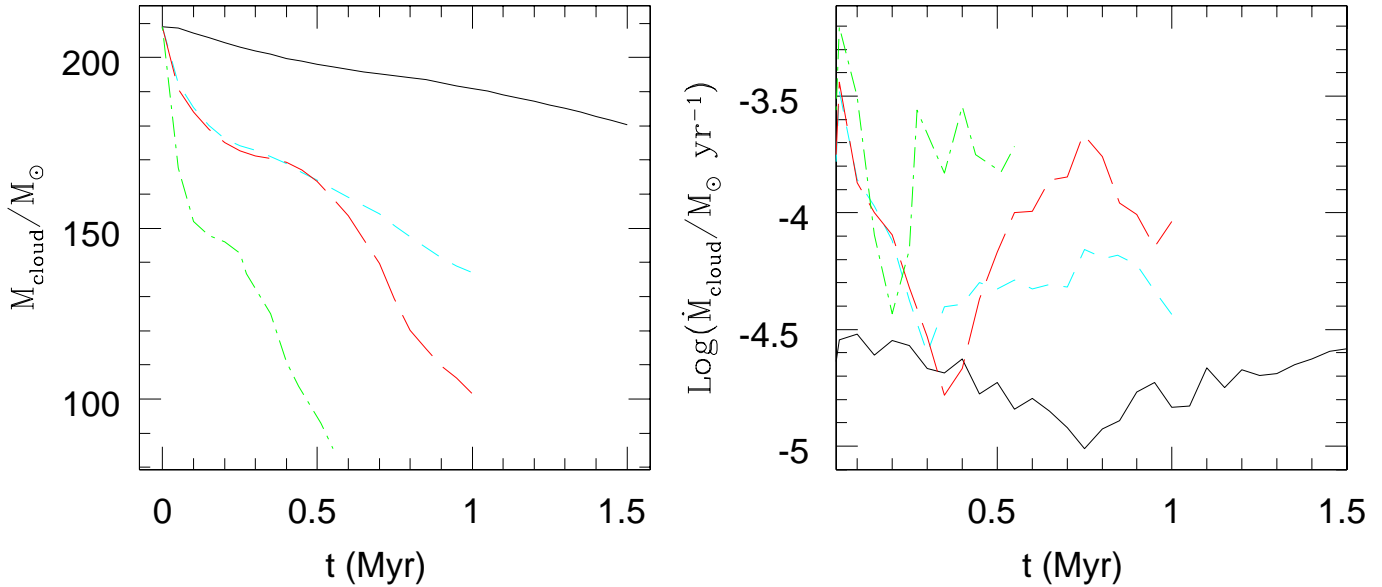}   
\end{center}   
\caption{Mass evolution (left panel) and mass loss rate (right panel) for   
the evaporating cloud models. Black solid lines : T1LP; blue short dashed: 
T1HP; red long dashed: T5LP; green dot dashed: T5HP.}  
\label{fig:massa}  
\end{figure*}

\emph{It is worth noting that, contrary to the same model without  
conduction, the cloud does not fragment because the R-T and K-H  
instabilities are strongly inhibited by the thermal conduction.}  
The K-H instability depends on the the steepness of  
the density and velocity gradients across the cloud surface. Since  
heat conduction smoothes out these gradients, the growth of the  
instabilities is reduced \citep{vieser00}. Furthermore  
the compression of the cloud due to the heat conduction  
reduces the momentum transfer from the superwind to the cloud, and  
thus the cloud acceleration responsible of the R-T instabilities.  
 
In Fig.~\ref{fig:massa} the evolution of the mass of the cloud and the
evaporation rate $\dot M$ are plotted for all the models with
conduction. This rate is measured by calculating the remaining cold gas
mass at each simulation time step, specifically the mass of gas with
$T < 1.2\times10^{4}$ K.  In the non-conductive models the cloud is
always heavily shredded, which makes it difficult to measure cloud
mass loss, but relatively little cold gas gets heated to $T >
1.2\times10^{4}$ K.  Focusing on model T5LP, at the beginning the
value of the mass loss rate is in agreement with
equation~\ref{equ:mdot_saturated} once the temperature and density of
the superwind behind the bow-shock is taken into account. As expected,
the rate drops by a factor of 20 after $t=0.35$ Myr, consistently with
the analogous reduction of the evaporating surface area proportional
to $\rc^2$. Then it increases again following an oscillatory behaviour
in pace with the evolution of the cloud surface. The reduction of
$\dot M$ is reflected in the evolution of the cloud mass which shows a
``flex'' in its temporal profile corresponding to the minimum of
the mass loss rate. The cloud lifetime turns out to be 2-3 times
longer than that provided by the constant rate $\dot M$ given by
equation~\ref{equ:mdot_saturated}.
 
The behaviour of all the other models is qualitatively similar to that  
of the reference model. The right panel in Fig.~\ref{fig:massa}  
reveals that the evaporative mass loss rate is initially higher in  
models with larger values of $v_{\rm w}$. The same models also show a  
larger frequency in the oscillatory behaviour of $\dot M$.  
Larger values of $v_{\rm w}$ lead to larger values of $T_{\rm  
bs}$ and of the heat flux. The larger the heat flux is, the faster the  
transmitted shock moves into the cloud; as a consequence, the  
re-expansion occurs earlier in models with larger $v_{\rm w}$. The  
similarity of the mass loss history for models with the same $v_{\rm  
w}$ (at least as long as the clouds maintain a rough spherical shape)  
is rather striking when one compares the temporal profiles of $M_{\rm  
c}$ for models T5LP and T1HP (see the left panel of  
Fig.~\ref{fig:massa}).

\subsection{X-ray and O{\sc vi} emission}  
\label{sec:results:emission}  
  
\subsubsection{O{\sc vi} and X-ray emission in NC models}  
\label{sec:results:emission:nocond}  
  
\begin{figure*}   
\begin{center}   
\psfig{figure=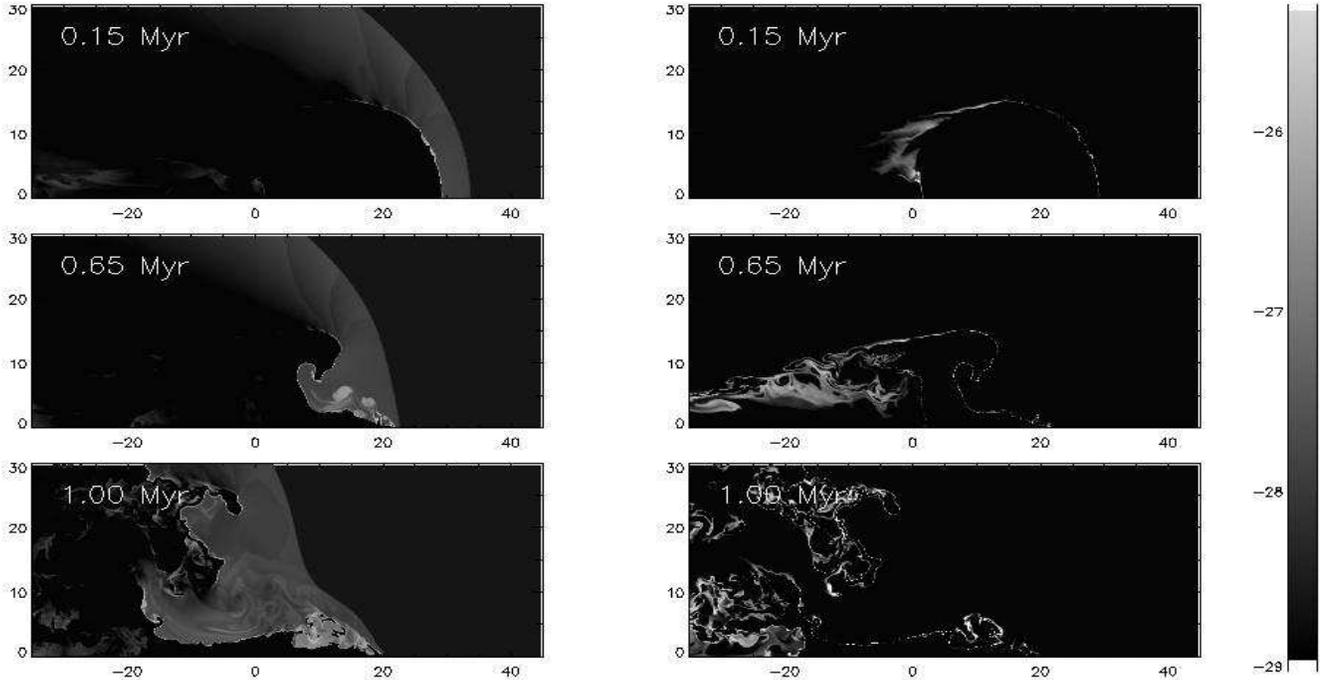}  
\end{center}   
\caption{X-ray (left panels) and O{\sc vi} (right panels) volume emissivity  
for the reference model T5LPNC at the same times of  
Fig.~\ref{fig:ral55}.  The logarithm of the soft X-ray (0.3 -- 2.0 keV  
energy band) and O{\sc vi} volume emissivity (in units of $\ergps  
\pcc$) is shown.  
}  
\label{fig:ral55emi}  
\end{figure*}

As pointed out in Section \ref{sec:results:static} when discussing  
clouds in a static hot medium, the interface between the cold  
dense cloud and the hot tenuous wind material is the source of both  
O{\sc vi} and soft X-ray emission. In the simulations without heat  
conduction these two phases tend to mix at the contact surface by  
numerical diffusion (see Fig.~\ref{fig:ral55emi}). While such  
diffusion can mimic qualitatively real phenomena like molecular  
diffusion and turbulent mixing, the quantitative effect can be easily  
overestimated giving rise to excessive radiative losses and  
overly-rapid cloud destruction.

It has been argued that the numerical spread of  
the contact discontinuities could in principle simulate, at least  
qualitatively, the presence of a heat conduction front; such an  
analogy, however, may be grossly in error in all the cases in which  
the conduction front is expected to give rise to dynamical effects  
such those described in the previous section, and which can not be  
generated by numerical diffusion.  As shown in the previous 
section  the inclusion of the thermal conduction significantly 
alters the dynamical behavior of the cloud. We also find that 
the radiative characteristics of the cloud change significantly.  
  
\begin{figure*}   
\begin{center}   
\psfig{figure=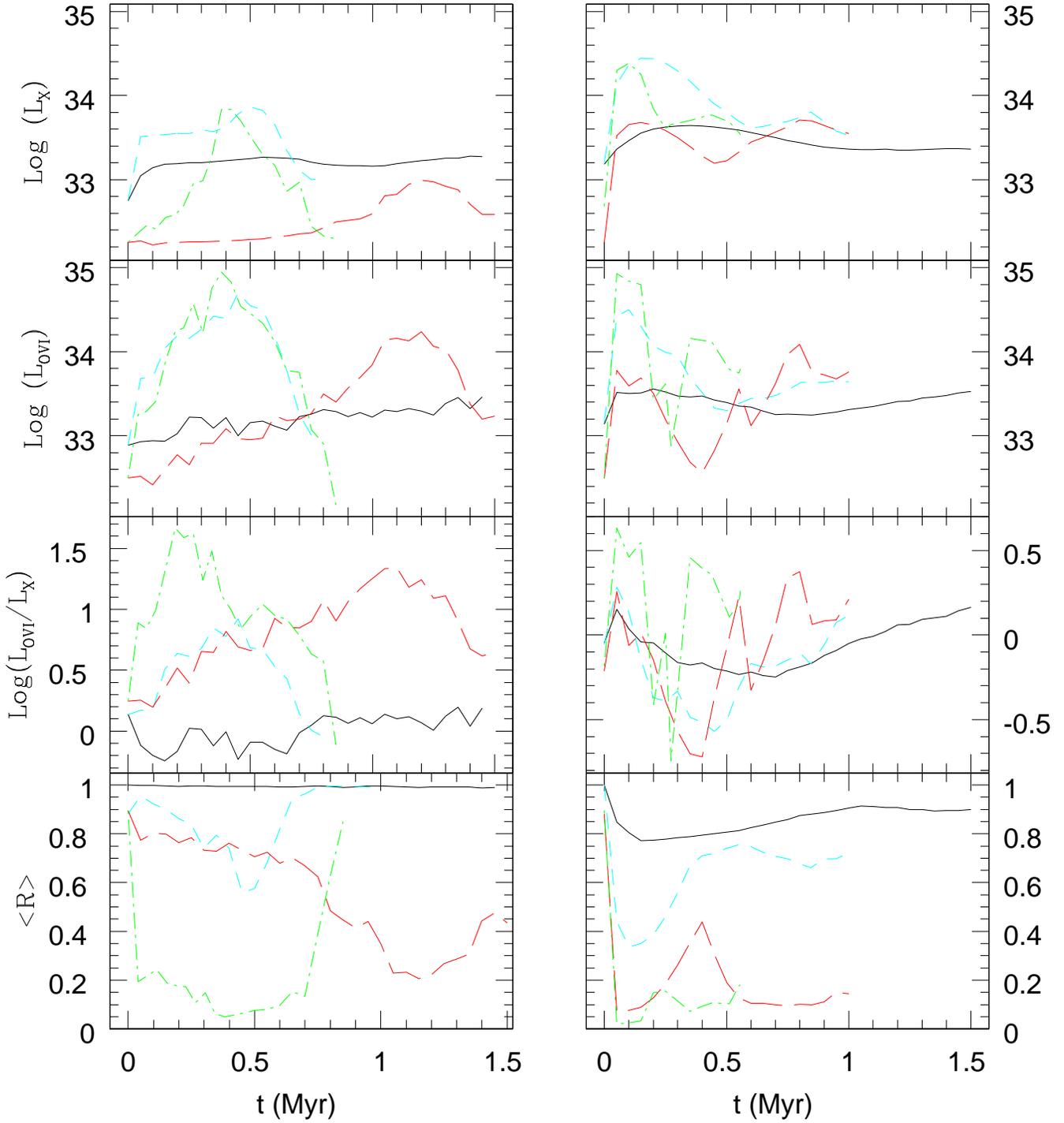}   
\end{center}  
\caption{Time evolution of the logarithm of X-ray (top panel)   
O{\sc vi} (second panel down) luminosities and the logarithm of their 
ratio (third panel down) and X-ray luminosity-weighted wind mass 
fraction $<{\cal R}>$ for all models. The right panels refer to models 
with the inclusion of effects due to heat conduction, while left 
panels refer to non-evaporative models NC. Black solid lines: T1LP; blue 
short dashed: T1HP; red long dashed: T5LP; green dot dashed: T5HP.} 
\label{fig:emi}  
\end{figure*}

In Fig.~\ref{fig:ral55emi} we plot the evolution of the O{\sc vi} and  
X-ray volume emissivity  (\ie luminosity per unit  
volume, $n_{\rm e} n_{\rm H} \Lambda(T)$) as a function of   
time for the reference model T5LPNC.   
In the left hand side panels of   
Fig.~\ref{fig:emi} the O{\sc vi} and X-ray luminosities, 
and X-ray luminosity-weighted wind mass fraction $<{\cal R}>$ of all the  
non-conductive models are compared.

The O{\sc vi} luminosity in model T5LP  
increases exponentially during the first $\sim 2.4  
\tau_{\rm cc}\sim 0.8$ Myr (see Fig.~\ref{fig:emi}), when the cloud,  
although highly distorted, is not yet fragmented. The rise in  
$\lo$ is due to the increase of the emitting surface area of the cloud.  
Subsequently the increase in $\lo$ is even more substantial as  
cloud fragmentation increases the cumulative emitting surface. Finally,  
after $t \sim 1.2$ Myr, the   
$\lo$ measured within the simulation volume  
drops as the cloudlets (which are more easily accelerated by the wind)  
start to leave the computational grid.

\begin{figure*}   
\begin{center}   
\psfig{figure=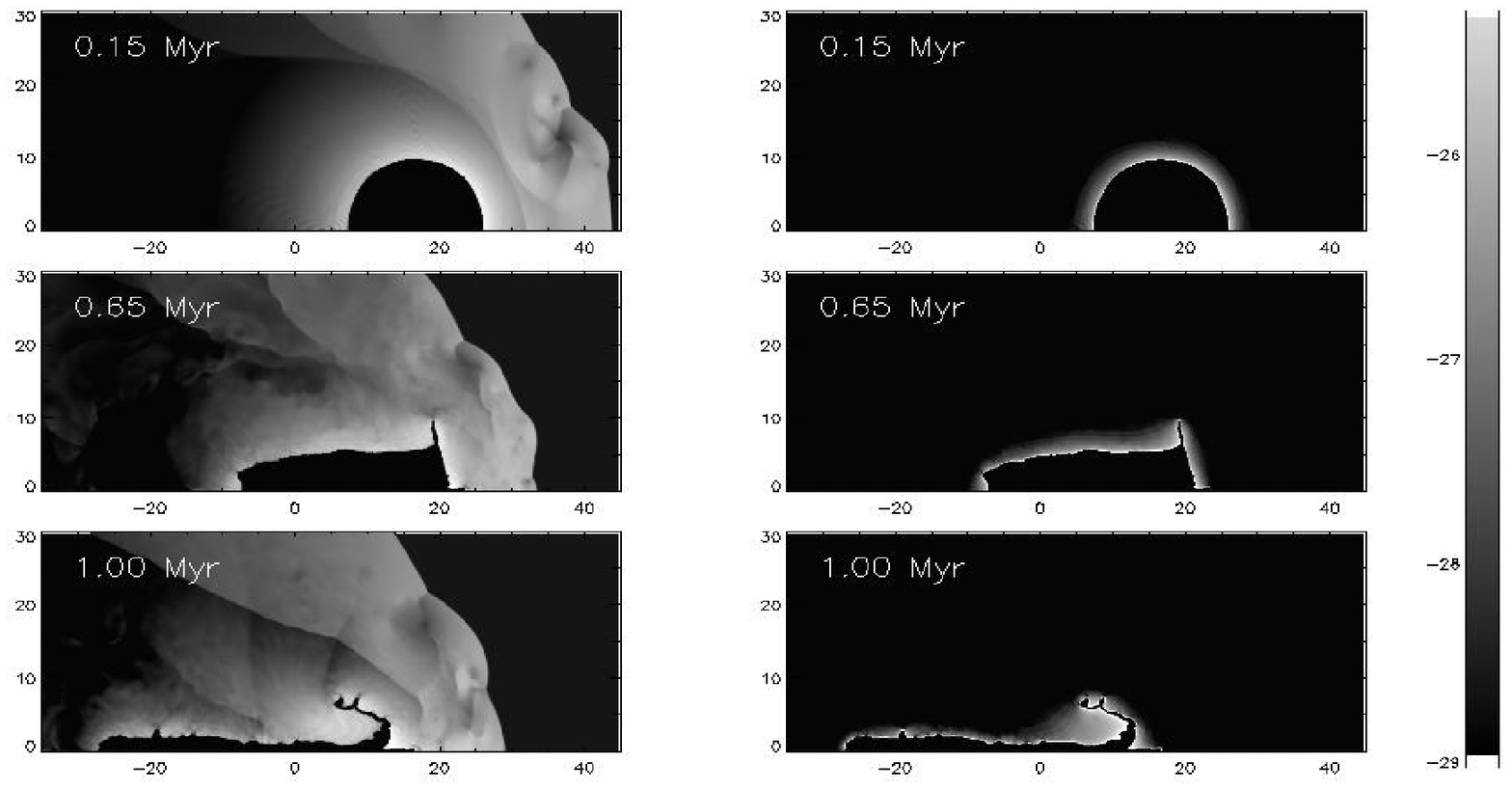}  
\end{center}   
\caption{X-ray (left panels) and O{\sc vi} (right panels) volume emissivity  
for the reference model T5LP at the same times of  
Fig.~\ref{fig:rar55}.  The logarithm of the soft X-ray (0.3 -- 2.0 keV  
energy band) and O{\sc vi} volume emissivity (in units of $\ergps  
\pcc$) is shown.   
}  
\label{fig:rar55emi}  
\end{figure*}   
   
Soft X-ray emission originates not only at the cloudlets' surfaces but  
also in the region behind the bow shock, where the temperature is  
$T\sim 1.6 \times 10^7$ K and the density $\sim 2.6 \times 10^{-3}$  
cm$^{-3}$.  
Wind-based material dominates the X-ray emission at times 
$t \la 0.8$ Myr, based on the X-ray luminosity weighted  
wind mass fraction ${\cal R}$ (bottom left panel of Fig.~\ref{fig:emi}). 
Once fragmentation becomes significant at $t \ga 0.8$ Myr the 
fraction of emission from cloud material increases to a maximum 
of $\sim 60$\%. 
It is worth noting that in Fig.~\ref{fig:emi} the  
contribution of the X-ray background due to the unperturbed hot  
superwind is also present. This contribution is of the same order of  
magnitude of that of the cloud, and this explains the rather low  
increase of $\lx$ during the first 0.8 Myr. Later, analogous to  
case of $\lo$, the X-ray luminosity shows a rise and fall  
due to cloud fragmentation and loss of cloudlets from the computational  
grid.

Model T1LPNC has the same ram pressure and the same analytical
crushing timescale $\tau_{\rm cc}$ of the model T5LPNC just discussed,
but lower superwind velocity and lower cloud to wind density ratio of
$\chi=100$. As pointed out by \citet{klein94}, the time at which the
cloud substantially fragments is also inversely proportional to
$\chi$.  This is confirmed by our simulations in which the model
T1LPNC fragments less dramatically than the model T5LPNC. This
different dynamical behaviour is responsible for the different time
evolution of $\lo$ in model T1LPNC, which increases at lower rate and,
in particular, does not show the "bump" present in the reference model
associated to the occurrence of the strong fragmentation. Note that
the initial value of $\lo$ in model T1LPNC is larger than in model
T5LPNC because of the different background contribution by the
superwind.  The X-ray luminosity in model T1LPNC is always dominated
by the emission from wind material in the region behind the bow shock
(${\cal R}$ is always $\sim 1$). Here the gas has a temperature $T\sim
3 \times 10^6$ K and a number density of $\sim 1.3 \times 10^{-2}$
cm$^{-3}$.  The density is nearly five times larger than in the
reference model, and thus $\lx$ is also larger.
   
Going back to the high ram pressure models T1HPNC and T5HPNC,
qualitatively the temporal evolution of $\lo$ and $\lx$ is similar to
the low pressure models, although occurring on shorter timescales.
This is as expected as for both T1HPNC and T5HPNC $\tau_{\rm cc}=1.5
\times 10^5$ Myr, shorter than in the models with lower ram
pressure. The increase in $\lo$ occurs at earlier times because of the
faster fragmentation, and is higher in magnitude in model T5HPNC
because of its higher value of $\chi$, as discussed above.  The strong
luminosity drop occurring at $t=0.5$ Myr is again due to the fact that
the cloud fragments leave the computational grid.  The luminosity peak
due to the fragmentation is also present in the X-ray luminosities,
and is associated with a drop in the fraction of the emission due to
wind material. Given the larger superwind density in model T1HPNC the
bow-shock is of relatively greater importance in this model, and the
relative rise in $\lx$ associated with cloud fragmentation is smaller
than in T5LPNC or T5HPNC.

\subsubsection{O{\sc vi} and X-ray emission of evaporating clouds  
dragged by the superwind}  
\label{sec:results:emission:cond}  
  
Fig.~\ref{fig:rar55emi} shows the O{\sc vi} and X-ray emissivity maps  
of the reference model T5LP at different times. As in the models  
without heat conduction, the X-ray flux originates not only close to  
the cloud surface, but also behind the bow-shock.   
The X-ray and O{\sc vi} volume emissivity is  
largest at the leading edge of the cloud, close  
to the symmetry axis, where the density of the evaporating gas has  
a maximum. This is due in part to the larger temperature of the shocked  
superwind, and partially to the compression given by the ram pressure.

In Fig.~\ref{fig:emi} (right panels) we plot the time evolution of  
$\lx$, $\lo$, $\lo/\lx$ and ${\cal R}$ for all the models with thermal  
conduction. Concentrating  
in particular on model T5LP, the O{\sc vi} and soft X-ray  
luminosities follow the same oscillating  
behaviour seen in mass loss rates of the cloud (Fig.~\ref{fig:massa}),  
and are similar to the analogous model at rest (see e.g.  
Fig.~\ref{fig:lo_lx_remodels}). Note that now the luminosities  
are systematically higher than in model RE05, and are more similar to  
those in model RE10. This is due to the  
larger evaporating flux of emitting gas due to the higher   
bow shock temperatures, as outlined above. The low values of  
${\cal R}$ demonstrate that the X-ray emission is now dominated 
by cloud material at essentially all times 
(at least in this case where $Z_{\rm w} = Z_{\rm c}$) 
When compared with the T5LPNC model (left hand side panels of  
Fig.~\ref{fig:emi}), both $\lx$ and $\lo$ are, at least at early  
times, up to 15-25 times higher than the values of the same model  
without conduction.  
  
The  $\lo/\lx$ ratio in these models is lower than  
in either the static conductive models, or in the non-conductive  
wind models. This is due to the combined effect of the heat  
conduction and of the bow shock, which tends to increase the X-ray flux  
more than the O{\sc vi} flux.  
It is only in model T1LP that conduction does not significantly 
increase the X-ray emission from the wind/cloud interaction. 
In all other conductive models X-ray luminosities are higher, 
and a larger and often dominant fraction of the emission comes 
from material that was originally part of the cloud.

The time evolution of the luminosities of the other conductive models is  
similar to that of the reference model T5LP (see Fig.~\ref{fig:emi}), and  
the relative differences in the oscillating period grossly follow the  
analogous differences found in the behaviour of $\dot M$ (see  
Fig.~\ref{fig:massa}).  In general, the amplitudes of the luminosity  
fluctuations are larger for models with larger values   
of $T_{\rm bs}$, and  
thus larger evaporative $\dot M$. 
 Note, however, that the luminosities in model T1HP  
reach higher values than in model T5LP although this latter develops  
essentially the same mass loss in the first 0.5 Myr. This is a  
consequence of the larger value of the ram pressure which reduces the  
volume between the cloud and the bow shock.  This is the region where  
the volume emissivities are largest,  
and where the evaporating gas in the front  
edge of the cloud moves before to be dragged downstream. The increase in   
the mean  
density of this region due to the reduced volume   
behind the bow shock thus leads to higher luminosities.

  
\subsection{O{\sc VI} absorption line column densities and kinematics}  
\label{sec:results:absorption}  
  
In addition to calculating the X-ray and O{\sc vi} emission from these  
models we also consider absorption line properties, specifically column  
densities of the O{\sc vi} ion probed in FUSE observations of  
starbursts \citep[see \eg][]{heckman01,heckman02}.  
  
Most of the physical and dynamical effects that alter the   
previously-discussed X-ray and O{\sc vi} emission also   
affect the gas that  
would be seen with absorption line probes. We have therefore kept this  
discussion short, and present only the time-averaged O{\sc vi}   
column densities for each of the models in  
Table ~\ref{tab:column_density}. For each model we report the  
mean value of $N_{\rm O VI}$ through the center of the cloud  
averaged over two different radii: $R_{\rm sor}=5$ pc   
and $R_{\rm sor}=15$ pc (this latter being equal to the cloud radius).   
Note that these radii give rise to a spatial area lower than that 
subtended by a FUSE LWRS aperture.

\begin{table*}   
\centering   
\begin{minipage}{170mm}   
\caption{Mean fractions of wind mechanical energy radiated   
and $\lo/\lx$ ratios in the wind/cloud interaction, and in the undisturbed 
wind.}  
\label{tab:percentual_luminosity}   
\begin{tabular} {lccccllr}  
\hline  
 Model & $L_{\rm mec}$ & $<\lx/L_{\rm mec}>$ & $<\lo/L_{\rm mec}>$ & $<L_{\rm tot}/L_{\rm mec}>$ &  \multicolumn{2}{c}{$<\lo/\lx>$} & $t_{\rm avg}$ \\  
      & erg $s^{-1}$ & (\%) & (\%) & (\%)  
	& ($Z_{\rm w}=Z_{\odot}$) &  ($Z_{\rm w}=10 Z_{\odot}$) & Myr \\  
(1) & (2) & (3) & (4) & (5)  
	& (6) & (7) & (8) \\  
 \hline   
%
T1LP(NC) & $1.34\times10^{36}$ & 0.23 (0.122) & 0.20 (0.13)   
	& 0.89 (0.55) & 0.87 (1.07) & 0.1 (0.11) 
	& 1.5  (1.5) \\  
	& \ldots & 0.045 & 0.028 & 0.138 & 0.62 & 0.62  
	& \ldots \\ 
T1HP(NC) & $1.50\times10^{37}$ & 0.07 (0.002) & 0.05 (0.11)   
	& 0.25 (0.37) & 0.71 (44.5) & 0.12 (5.23) 
	& 1.0  (0.8) \\  
	& \ldots & 0.0040 & 0.0025 & 0.0123 & 0.62 & 0.62 
	& \ldots \\ 
T5LP(NC) & $3.00\times10^{36}$ & 0.12 (0.013) & 0.13 (0.14)   
	& 0.55 (0.49) & 1.11 (11.2) & 0.34 (1.68) 
	& 1.0  (1.5) \\  
	& \ldots & 0.0062 & $9.7\times10^{-7}$ & 0.0062 &  
	$1.6\times10^{-4}$ & $1.6\times10^{-4}$ 
	& \ldots \\ 
T5HP(NC) & $3.33\times10^{37}$ & 0.03 (0.004) & 0.07 (0.08)   
	& 0.29 (0.26) & 2.68 (18.5) & 1.31 (7.58) 
	& 0.55 (0.8) \\  
	& \ldots & 0.00056 & $8.8\times10^{-8}$ & 0.00056 & 
	$1.6\times10^{-4}$ & $1.6\times10^{-4}$ 
	& \ldots \\ 
 \hline  
 \end{tabular}  
\par  
\medskip  
Notes --- The time-averaged mean X-ray, O{\sc vi} and total luminosities are  
given as a fraction of the superwind mechanical power $L_{\rm mec} =   
0.5 \, \rho_{\rm w} \, v_{\rm w}^{3} \, \pi \, R^{2}_{\rm em}$  
passing through the  
chosen volume (a cylinder of radius $R=10$ pc and extending 20 pc   
upstream and 80 pc downstream of the initial cloud center).  
Note that the luminosities are calculated over a larger volume, with  
a radius of $50$ pc. See Section \ref{sec:discussion:luminosities} for  
details.  
Columns are as follows: Column 1: Model name. Values given   
in parentheses correspond the cloud models without thermal conduction.  
The second line of values associated with each model shows the 
$\lx/L_{\rm mec}$, $\lo/L_{\rm mec}$, $L_{\rm tot}/L_{\rm mec}$ 
and $\lo/\lx$ values for an equal volume of wind material with 
no clouds within it.  
Column 2: Wind mechanical luminosity $L_{\rm mec}$.  
Column 3: Average soft X-ray  
emission (E=0.3-2.0 keV) to wind mechanical power ratio (in percent). 
All values in columns 3 -- 6 were calculated using  
$Z_{\rm w} = Z_{\rm c} = Z_{\odot}$. 
Column 4: Average O{\sc vi} doublet ($\lambda=1032$ \AA~and 1038 \AA)   
emission to wind mechanical power ratio (in percent). Column 5:  
Average total FUV and soft X-ray emission  
to wind mechanical power ratio (in percent), assuming the   
O{\rm vi} emission accounts  
for 30\% off the total FUV cooling rate \citep{hoopes03}. Column   
6: Average O{\sc vi} to soft X-ray luminosity ratio.  
Column 7: As column 6, except calculated with $Z_{\rm c} = Z_{\odot}$ and 
$Z_{\rm w} = 10 Z_{\odot}$. 
Column 8:   
The time over which the luminosities were  
averaged.   
\end{minipage}   
\end{table*}   
  

Thus, if the areal covering fraction of clouds within an LWRS is  
of order unity, then this is the column density that would be observed.  
But, if the cloud covering factor is less, then the column  
density that would be observed with the LWRS on FUSE would  
be proportionally less.  
  
In the models without conduction the O{\sc vi} column density $N_{\rm
O VI}$ varies with time, depending on the fragmentation of the cloud,
and roughly proportional to the behaviour of $\lo$.  From
Table~\ref{tab:column_density} it is clear that the time-averaged mean
column density $<N_{\rm O VI}>$ is higher for models with larger wind
ram pressure (and consequently shorter cloud crushing time $\tau_{\rm
cc}$). Models with the same ram pressure have roughly the same value
of $N_{\rm O VI}$ (within 0.1 dex).  Analogously to our earlier
discussion on the $\lo$ behaviour, the model T5LPNC has a slightly higher
mean $N_{\rm O VI}$ value compared to equivalent ram pressure model
T1LPNC, most probably because of the higher cloud fragmentation (which
depends on the cloud density contrast $\chi$).  An effect of similar
magnitude is seen comparing the non-conductive high pressure models.
   
These effects are most clearly seen along lines of sight weighted
toward the center of the original cloud where the density of
$T\sim10^{5.5}$ K gas is greatest, \ie $R_{\rm sor}=5$ pc in
Table~\ref{tab:column_density}. Mean density columns over a larger
radius still show the same general pattern in these non-conductive
models, as cool clouds fragment and their O{\sc vi} interfaces extend
to radii greater than that of the original cloud.
  
The inclusion of thermal conduction alters the mean O{\sc vi} columns
as follows. Conductive effects crush the cloud, and cloud
fragmentation is almost completely suppressed. As a result there is a
large difference between the mean $N_{\rm O VI}$ column density averaged
over a line of sight of 5 pc radius (which always intersects the
cloud) to that averaged over 15 pc radius (which often includes line
of sight with negligible O{\sc vi}-absorbing material).

Considering $R_{\rm sor}=5$ pc, the $<N_{\rm O VI}>$ value observed is
physically driven by the bow shock temperature $T_{\rm bs}$ (as is the
case for $\lo$), with higher $T_{\rm bs}$ driving more conductive
evaporation and hence creating more $T\sim 10^{5.5}$ K gas. Note how
much closer the $N_{\rm O VI}$ values are between models T1HP and T5LP
(which have very similar post-shock temperatures $T_{\rm bs}$) than in
the case of the equivalent non-conductive models. The higher $N_{\rm
O VI}$ value for T5LP is due to the higher initial temperature of the
wind in T5LP. This hotter gas also drives stronger evaporation behind
the cloud (on the reverse side to the wind bow shock) than in model
T1HP.
  
When averaging the O{\sc vi} column over the larger radius of $R_{\rm
sor}=15$ pc, the time-averaged $N_{\rm O VI}$ is relatively uniform,
even slightly decreasing as $T_{\rm bs}$ increases. Here the strong
increase in $N_{\rm O VI}$ with increased $T_{\rm bs}$ along lines of
sight passing through the cloud is offset by increased numbers of
lines of sight with negligible $N_{\rm O VI}$, as increased $T_{\rm
BS}$ leads to stronger cloud compression and reduces its
cross-section.
 
Within the short period of time over which we can follow the evolution
of the cloud in these simulations the cold gas is accelerated by the
wind to a mass-weighted average velocity between 20 and $100$ km
s$^{-1}$ (at $t=0.5$ Myr). Cold gas velocities are systematically
larger in the high ram pressure models compared to low pressure
models. Comparing equivalent conductive and non-conductive models, the
mean cold gas velocity is higher in the non-conductive models as cloud
fragments are easily accelerated to higher velocity.  The fraction of
the wind mechanical energy converted to cloud kinetic energy by this
time (calculated within a radius of 10 pc for all models) ranges
between 0.6 and 6\%, and on average is 2\% for models with conduction
and 5\% for non-conductive models. Note that more wind mechanical
energy is converted to cloud kinetic energy than is radiated away (see
Table~\ref{tab:percentual_luminosity}).
 
We investigated the O{\sc vi} kinematics of a randomly-chosen subsample 
of the simulated O{\sc vi} absorption lines.  Mean velocities were 
in the range 0 to 200 km s$^{-1}$, although we did not find any clear 
correlation between mean cold gas velocities and O{\sc vi} profile velocities. 
We found line widths equivalent to $b$ values in the  
range 30 to 200 km s$^{-1}$.

\begin{table*}  
\centering   
\begin{minipage}{110mm}   
\caption{Mean O{\sc vi} column densities through the simulated clouds.}  
\label{tab:column_density}   
\begin{tabular} {|lcclcc}   
\hline  
Model & \multicolumn{2}{c}{log ($N_{\rm O VI}$)} & Model & \multicolumn{2}{c}{log ($N_{\rm O VI}$)} \\  
\hline  
      & $R_{\rm sor}=5$ pc & $R_{\rm sor}=15$ pc &    & $R_{\rm sor}=5$ pc &  $R_{\rm sor}=15$ pc \\  
\hline   
%
%
%
%
T1LP   & 12.86    & 12.88    & T1LPNC  & 13.08    & 12.54 \\  
T1HP   & 13.29    & 12.83    & T1HPNC  & 13.29    & 12.98 \\  
T5LP   & 13.43    & 12.80    & T5LPNC  & 13.19    & 12.97 \\  
T5HP   & 13.52    & 12.80    & T5HPNC  & 13.38    & 13.05 \\  
\hline  
\end{tabular}  
\par  
\medskip  
Notes --- Time-averaged mean values of $\log N_{\rm O VI}$   
(in units of $cm^{-2}$) due to the wind/cloud  
interaction in each of the models. See Section \ref{sec:results:absorption}  
for details. The O{\sc vi} column density due to the   
undisturbed wind within the computational volume  
has been evaluated and subtracted from the   
total derived O{\sc vi} column density, \ie the quoted column   
density is the additional O{\sc vi} column generated by the   
wind/cloud interaction. The undisturbed wind has  
 $\log N_{\rm O VI}=12.65$ in all T1 models, and $\log N_{\rm O VI}=10.04$  
in all T5 models.   
\end{minipage}   
\end{table*}

\subsection{The influence of the cloud radius}  
\label{sec:results:cloud_radius}  
  
In all the previously described simulations, the initial radius of the
cool cloud was 15 pc, consistent with the smallest scale structures
observed in the warm ionized gas in superwinds and seen in numerical
simulations of superwinds (as discussed in Section
\ref{sec:introduction}).  This does not mean that cool clouds of only
this size exist.  Much larger structures of relatively cool $T\la
10^{4}$ K gas exist (\eg the northern cloud or cap in M82, see
\citealt{lehnert99}, or the large scale filaments in the halo of NGC 253)   
although how coherent these structures are is hard to assess, and
these very large structures may actually mark to edges of the flow
rather than being fully embedded within the wind.
 
The distribution of cloud sizes and masses in winds has not been
determined observationally or theoretically. As the average cloud size
may differ from the $\rc=15$ pc used as a default in our simulations
it is worth considering how the emission and absorption properties of
a superwind/cloud interaction depend on cloud radius.  To investigate
this we ran one further model, with a cloud of radius $\rc = 45$ pc
(three times larger than in all the previously discussed models), on a
grid with resolution $\Delta r = \Delta z = 0.3$ pc. All other model
parameters were the same as in the low pressure conductive model T1LP.

\begin{figure}   
\begin{center}   
\psfig{figure=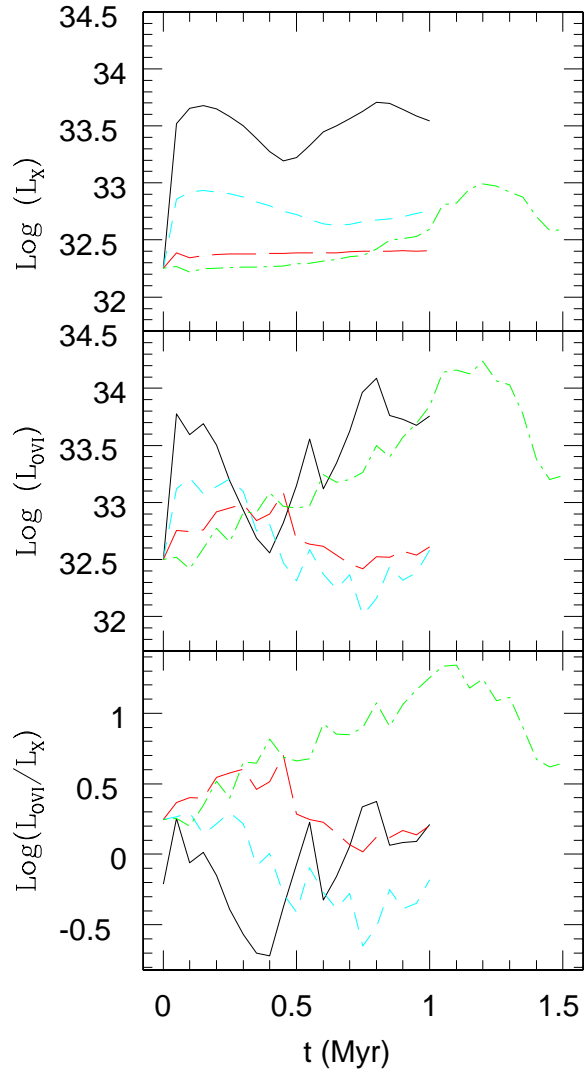,width=8.5cm} 
\end{center}   
\caption{Time evolution of the logarithm of the X-ray (upper panel),  
O{\sc vi} (middle panel) and the logarithm of their ratio (bottom 
panel) for model T5LP for different values of the reduction factor $f$ 
of the Spitzer heat conduction coefficient. Balck solid line: $f=1$; blue 
short dashed: $f=5$; red long dashed: $f=25$; green dot dashed: $f=\infty$.} 
\label{fig:reduction}  
\end{figure}

We compared the soft X-ray luminosity, O{\sc vi} luminosity and
time-averaged O{\sc vi} column density in this large radius model to
the values found in model T1LP, after subtracting the background due
to the undisturbed superwind within the same volume.  As described in
Section \ref{sec:numerical_method}, for the model with $\rc=15$ pc the soft
X-ray and O{\sc vi} emission was calculated within a radius of 50 pc
off the cloud axis, and 20 pc upstream and 80 pc downstream of the
initial location of the cloud. As this is similar to the size of the
cloud in the larger cloud radius model we also calculated the
\emph{emission} from the large cloud radius model within a region 3
times larger in each dimension than that used in the default
models. Column densities were calculated within the same 5 and 15 pc
radii, but along a path length 3 times larger.  As the size of regions
used differ between the default model and the large cloud radius
model, we subtracted the emission or absorption contribution
expected from the undisturbed wind within the region to allow a fair,
if approximate, comparison of the relative change in luminosity or
column density.
  
The soft X-ray emission in the large cloud radius model is $\sim 40$  
times greater than in model T1LP, \ie scaling roughly as   
$L_{\rm X} \propto \rc^{3.3}$. As the soft X-ray emission in this  
supersonic wind/cloud interaction mainly arises in the bow  
shock (in model T1LP conduction plays less of a role in enhancing the  
X-ray emission than in the other conductive model), this   
scaling can be understood as cross section of the bow shock scaling  
as $\rc^{2}$ and its thickness scaling as $\rc^{1}$, and thus the luminosity  
scales as the approximate increase in shocked gas volume.  
  
In contrast to the X-ray emission, which is spread over a large volume within  
the bow shock, the O{\sc vi} emission is concentrated onto a relatively  
thin interface region at the immediate edge of the cloud core.   
The relative increase in $\lo$ is only a factor $\sim 12$,   
which indicates that $\lo \propto \rc^{\sim 2}$. This suggests that  
the thickness of O{\sc vi} zone of the conductive front is relatively  
independent of the cloud radius (to within $\sim 30$\%), and that  
most of the O{\sc vi}   
luminosity increase is due to increased cloud surface area.  
  
This conclusion is supported by the simulated O{\sc vi} column
densities, which are very similar between the large cloud radius model
and model T1LP. For the large cloud radius model the time averaged
mean is $\log N_{\rm O VI} = 12.89$ for $R_{\rm sor}=5$ pc (higher by
$\sim 7$\%), and $\log N_{\rm O VI} = 13.11$ for $R_{\rm sor}=15$ pc
(higher by $\sim 70$\%).  As in model T1LP the cloud is compressed to
radii less than 15 pc so that some lines of sight see little or no
O{\sc vi}-absorbing gas, the fairest comparison is between the columns
averaged over the $R=5$ synthetic beam. The similarity in column
density is thus consistent with a conductive interface whose density
and thickness is independent of the cloud radius to first order.
  
Thus these results indicate that important emission properties,  
specifically the soft X-ray emission, but also the related   
$\lo/\lx$ ratio and the fraction of wind energy lost to  
radiation ($\lx/L_{\rm mec}$ and $L_{\rm tot}/L_{\rm mec}$), do depend  
on the cloud radius (larger clouds will give lower $\lo/\lx$ ratios and  
higher $L_{\rm X}/L_{\rm mec}$ ratios).

\subsection{Lower values of the coefficient of thermal conductivity}  
\label{sec:results:reduced_conductivity} 
 
Both X-ray observations of the hot inter-cluster medium 
\citep[\eg][]{ettori00,vikhlinin01} and theoretical  
considerations \citep*{narayan01,asai04,chandran04},  
suggest that the actual heat conduction  
coefficient is reduced in the presence of tangled magnetic fields  
by a factor of at least 5 relative to the Spitzer value. 
In order to understand how sensitive our results are to the absolute 
value of the conductivity coefficient $\kappa$ we ran two more  
models analogous to model T5LP, but reducing this coefficient by 
factors 
$f=5$ and $f=25$ respectively,  
i.e. $\kappa=\kappa_{\rm Sp}/f$, where $\kappa_{\rm Sp}=6.1\times  
10^{-7}T^{5/2}$ erg s K$^{-1}$ cm$^{-1}$ is the ``standard" Spitzer  
value (adopted in Eq. 1 and in all the simulations with heat  
conduction discussed so far. Thus $f=1$ represents model T5LP, and  
$f=\infty$ is model T5LPNC).  
We also ran a model with $f=125$, but the results were very similar to  
those with $f=\infty$, so we do not discuss this model further. 
 
As expected the apparent strength of the R-T and K-H instabilities  
increase as $f$ increases. However, within $t=1$ Myr (the time  
at which the simulations have been stopped) the cloud does not  
undergo any significant fragmentation even for $f=25$. This fact  
influences the temporal behaviour of $L_{\rm X}$ and $L_{\rm OVI}$. The  
upper panel of Fig.~\ref{fig:reduction}  
illustrates $L_{\rm X}$ temporal profiles of the  
model T5LP for the four values of $f$, $f=(1,5,25,\infty)$. 
The profile for $f=5$ is similar in shape 
to that for $f=1$ but at lower X-ray luminosity because of the weaker 
conductive mass loading of the bow shock region. 
The oscillations are weaker because the  
reduced heat conduction produces smaller oscillations of the cloud  
radius. The X-ray luminosity for $f=25$ is rather constant  
and only slightly larger than that for $f=\infty$. Note however that,  
for $t> 0.8$ Myr, $L_{\rm X}$ for the model T5LPNC increases because  
of the fragmentation, while remains nearly constant for $f=25$. For  
the same reason, $L_{\rm OVI}$ remains quite low for $f=5$ and $f=25$  
relative to $f=\infty$ (second panel of Fig.~\ref{fig:reduction}).  
 
The cloud lifetime for these models with  
conduction of intermediate strength ($5<f<25$) 
is increased because of the reduced evaporation rate while the 
fragmentation remains essentially inhibited.

\begin{table}  
\centering   
\begin{minipage}{\columnwidth}   
\caption{Mean O{\sc vi} to X-ray flux ratios and  
O{\sc vi} column densities for different values  
$f$ of the reduction of the coefficient of thermal conductivity, in  
models based on T5LP conditions. All calculations assume  
$Z_{\rm w} = Z_{\odot}$.}  
\label{tab:column_density_reducted}   
\begin{tabular} {lrll}   
\hline  
$f$ & $<L_{\rm O VI}/L_{\rm X}>$ & \multicolumn{2}{c}{log ($N_{\rm O VI}$)} \\  
\hline  
      & & $R_{\rm sor}=5$ pc & $R_{\rm sor}=15$ pc \\ 
\hline   
 
1   & 1.1 & 13.43  & 12.80  \\  
5   & 1.0 & 12.39  & 12.04  \\ 
25  & 2.3 & 12.11  & 12.20  \\          
$\infty$ & 11.2 & 13.19  & 12.97  \\  
 
\hline  
\end{tabular}  
\par  
\medskip  
Notes --- see Section \ref{sec:results:reduced_conductivity} for details. 
\end{minipage}   
\end{table}  

In Table~\ref{tab:column_density_reducted} we summarize the time-averaged 
mean ratio $L_{\rm OVI}/L_{\rm X}$ for different values of $f$. 
It is interesting to note that this 
value is essentially the same for values of $f=1$ and $f=5$, and 
only increases to 2.3 for $f=25$. 
As shown in Fig.~\ref{fig:reduction} the increase in 
$L_{\rm OVI}/L_{\rm X}$ ratios with increasing $f$  
is due to the gradual reduction in $L_{\rm X}$ 
while $L_{\rm OVI}$ values are relatively similar irrespective 
of the strength of thermal conduction.

It is also interesting to compare the different values of the mean OVI  
column density for different values of $f$ (shown in  
Table~\ref{tab:column_density_reducted}), where two  
values of $R_{\rm sor}$ are considered. For $R_{\rm sor}=15$ the  
column density is reduced by a factor of 5 as $f$ increases from $f=1$  
to $f=5$. This can be easily understood because the amount of the OVI  
emitting gas is roughly proportional to the rate of mass loss, and  
thus to $f^{-1}$. We point out that for $R_{\rm sor}=5$ the difference  
is larger because, contrary to the case $f=1$, with $f=5$ the cloud  
shrinks less and does not fit entirely inside $R_{\rm sor}$.  As $f$  
increases further, the column density reverts this trend and increases  
again. This is due to the effect of the numerical diffusion which  
starts to prevail on the thermal conduction in spreading the cloud  
edge when $5<f<25$. The column density increase for $f>25$ is due to  
the occurrence of the fragmentation and the subsequent increase of the  
emitting surface.

These results suggest that even with thermal conduction inhibited 
by realistic amounts the interesting dynamical and radiative effects 
of conduction on wind/cloud interactions are retained. X-ray emission 
is enhanced, the $L_{\rm OVI}/L_{\rm X}$ ratio remains low, and 
hydrodynamical cloud fragmentation is inhibited.

%
  
\section{Discussion and implications for superwinds}  
\label{sec:discussion}  
  
\begin{table*}  
\centering   
\begin{minipage}{165mm}   
\caption{Observed {\it FUSE} O{\sc vi} to soft X-ray flux ratios in 
superwinds and normal spiral halos.}  
\label{tab:obs_flux_ratios}   
\begin{tabular}{lrllllrr}  
\hline  
Galaxy and region & Width   
	& $f_{\rm O VI}$ & $f_{X, 0.3-2.0}$ & $\log L_{\rm X}$  
	& $f_{\rm O VI}/f_{\rm X}$   
	& O{\sc vi} ref. & X-ray ref. \\  
        & pc             & $\ergps \pcmsq$ & $\ergps \pcmsq$ &   
	& & \\   
(1)     & (2)  
	& (3) & (4) & (5) & (6) & (7) & (8) \\  
\hline  
M82-A   & 525 & $<11.4\times10^{-14}$ & $(94.6\pm{2.2})\times10^{-14}$ & 39.2 &  
	$<0.12$ & 1 & 1 \\   
M82-B   & 525 & $<11.4\times10^{-14}$ & $(69.5\pm{2.1})\times10^{-14}$ & 39.0 &  
	$<0.16$ & 1 & 1 \\   
M82-C   & 525 & $<9.3\times10^{-14}$ & $(11.9\pm{0.8})\times10^{-14}$ & 38.3 &  
	$<0.78$ & 1 & 1 \\   
M82-D   & 525 & $<9.5\times10^{-14}$ & $(6.3\pm{0.5})\times10^{-14}$ & 38.0 &  
	$<1.51$ & 1 & 1 \\ 
NGC 3079 & 2490 &   $<1.5\times10^{-14}$ & $(30.7\pm{2.9})\times10^{-14}$  
	& 40.0 & $<0.05$ & 2 & 2 \\  
NGC 4631-A   & 1092 & $(3.6\pm{0.8})\times10^{-15}$ & $(3.6\pm{0.9})\times10^{-15}$ & 37.4 &  
	$1.00\pm{0.33}$ & 3 & 4 \\   
NGC 4631-B   & 1092 & $(6.2\pm{1.0})\times10^{-15}$ & $(10.0\pm{1.2})\times10^{-15}$ & 37.8 &  
	$0.62\pm{0.12}$ & 3 & 4 \\   
NGC 891-2   & 1395 & $(2.1\pm{1.1})\times10^{-15}$ & $(7.8\pm{3.8})\times10^{-15}$ & 37.9 &  
	$0.27\pm{0.19}$ & 3 & 4 \\   
\hline  
\end{tabular}  
\par  
\medskip  
A compilation of observed FUSE O{\sc vi} flux measurements   
in the halos of nearby starburst and normal spiral galaxies,  
along with {\it Chandra} ACIS-S measurements of the diffuse  
soft X-ray luminosity within the same region. The physical size  
corresponding the $30\arcsec$ angular width of the square FUSE LWRS  
aperture is given in column 2, assuming distances of 3.6, 17.1, 
7.5 and 9.6 Mpc to M82, NGC 3079, NGC 4631 and NGC 891 respectively.  
The O{\sc vi} fluxes quoted in column 3   
refer to the sum of the 1032 and 1038~\AA~lines,   
assuming the flux in the 1038~\AA~line is exactly half that  
of the 1032~\AA~fluxes measured (the reference for which is  
given in column 7).  
The diffuse soft X-ray ($E=0.3-2.0$ keV)   
fluxes are based on {\it Chandra} ACIS-S  
measurements, the reference for which is given in column 8.  
The log of the absorption-corrected  
soft X-ray luminosity (in units of $\ergps$) within the FUSE  
observation region is given in column 5.   
Upper limits are $3\sigma$. Errors in fluxes, where given,   
are purely the statistical uncertainties in the number of  
O{\sc vi} 1032~\AA~or diffuse $E=0.3-2.0$ keV X-ray photons  
within the region encompassed by the FUSE aperture.  
All fluxes are corrected for extinction/absorption, using the  
Galactic extinction law \citep{cardelli89} for O{\sc vi} and  
the \citet{morrison83} model for X-ray absorption cross sections.   
Only Galactic foreground extinction is assumed  
for NGC 4631 and NGC 891, while intrinsic absorption within the  
wind is required for M82 \citep{hoopes03} and NGC 3079 (Hoopes et al, in 
preparation).  
\par  
\smallskip  
References: (1) \citet{hoopes03}. The X-ray fluxes are based on spectral  
fits to the diffuse X-ray emission within each of the regions covered by the  
FUSE apertures. (2) From Hoopes et al (in preparation). 
(3) \citet{otte03},   
(4) This work. X-ray fluxes are based on scaling the measured  
halo-region diffuse soft X-ray fluxes from \citet{strickland04a}  
by the ratio of the $E=0.3-2.0$ keV diffuse count rate in the  
region of the FUSE aperture to the total halo diffuse count rate.  
\end{minipage}  
\end{table*}

\subsection{Cloud survival in superwinds}  
\label{sec:dicsussion:cloudlife}  
  
In order to produce the broad absorption line profiles of  
warm, neutral and coronal gas observed in superwinds, ambient  
ISM material must survive being entrained within the hotter superwind  
material long enough for it to be accelerated from rest to velocities    
of several hundred km/s \citep{heckman2000,heckman01,heckman05}.  
Warm ionized gas is observed to extend out to heights of $H_{\rm WIM}   
\ga 10$ kpc   
from the host galaxy in many superwinds. If this material has  
been transported   
from an original location within the disk this would imply a life time   
of $t \sim 2 \, (H_{\rm WIM}/{\rm kpc})/(v_{\rm WIM}/500 \kmps)$ Myr.  
Yet clouds over-run by a blast wave or enveloped in a supersonic  
wind are expected to be hydrodynamically  
destroyed within a few crushing time scales \citep[see][]{klein94}, which  
are typically less than 1 Myr for typical estimates of superwind and cloud  
parameters (see Table~\ref{tab:cloud_in_wind_params}).   
  
Understanding the conditions most conducive to  
cloud survival is therefore important. Theoretical models   
in which clouds are rapidly destroyed are also less likely   
to be good representations  
of the physics of superwinds than those in which clouds have longer  
lifetimes.  
  
In this work we find that cloud models with thermal conduction   
have very different properties  
in terms of cloud survival to those without conduction.  
Thermal conduction suppresses the instabilities that strongly fragment the  
cloud in the non-conductive case, but also leads to significant  
cloud mass loss by  
evaporation into the hotter phases.  
  
The mass and mass loss rate in cold gas ($\log T < 4.08$) is   
shown as a function of time in Fig.~\ref{fig:massa} for   
the models with thermal conduction.  
The majority of this mass loss is due to conductive-driven  
evaporation, as cold cloud gas is heated  
to $T \ga 10^{5}$ K. As previously discussed in   
Section \ref{sec:results:evap_clouds}, the net evaporation rate is  
proportional to some power of the temperatures of the bow shock   
around the cloud, so the evaporation rate is least in the low  
wind velocity, low wind temperature model T1LP. The saturation 
of thermal conduction reduces the evaporation rates compared 
to the classical case (Equ.~\ref{equ:mdot_saturated}), but this analytical 
theory does not include dynamical effects (such as cloud compression) 
that further reduce the evaporation rate in models T1HP, T5LP and T5HP. 
The net effect is for an increase cloud survival times in cases where 
$\sigma_{0} \ga 0.25$ in comparison to the values predicted 
from \citet{cowie77}. 
 
If we define  
a cloud lifetime as $\tau_{\rm life} =   
M_{\rm cloud}/\Mdot_{\rm cloud}$, and 
using the mean cloud mass loss rate found in the simulations,   
then the cloud life times in the conductive  
models are as follows. Model T1LP: 12.8 Myr; Model T1HP: 3.3 Myr;  
Model T5LP: 2.0 Myr; Model T5HP: 0.9 Myr.  
With the exception of  
model T1LP, these lifetimes are somewhat less than the lifetime we  
expect the average cloud in a superwind should have.  
 
The mean mass loss rate in the simulation of the cloud with $\rc=45$ pc 
is $\Mdot_{\rm cloud} = 6.26\times10^{-5} M_{\odot} \pyr$,  
three times larger than in the 
analogous model T1LP with the standard radius $\rc=15$ pc  
($1.93\times10^{-5} M_{\odot} \pyr$). Conduction is not saturated 
in this model, as $\sigma_{0}=0.043$, so that the mass loss rate is 
classical and scales in proportion $\rc$. The numerical result is  
consistent with the theoretical expectation.  
In fact, the \citeauthor{cowie77} 
formula (Equ.~\ref{equ:mdotcl})  
yields $\Mdot_{\rm cloud} = 1.84\times10^{-5} M_{\odot} \pyr$  
for $\rc=15$ pc, and $5.52\times10^{-5} M_{\odot} \pyr$ for 
$\rc=45$ pc (These values are obtained considering a 
     temperature $T=2\times10^{6}$ K measured from the simulations  
     between the bow shock and the cloud [instead of the theoretical bow-shock 
temperature], and halving the values obtained considering 
     the lower temperature at the rear of the cloud). 
     The cloud lifetime scales as $R^2$ in the classical 
  regime, as expected. Note 
     however that if dynamical effects were important (as in all 
     models but T1LP) we expect 
     a lifetime longer than the theoretical one. 
To summarize, cloud lifetime is a strong function of cloud size. Conductive 
clouds larger than the default $\rc=15$ pc we consider may well have survival 
times equivalent to wind flow times, even in the case of conduction 
at Spitzer levels. 
  
It is difficult to assess cloud lifetimes in the non-conductive models. Cloud  
fragments are very rapidly accelerated and advected off the fixed numerical  
grid we have used in these simulations. It is hard to assess what fraction  
of the mass loss from the cool phase is due to this process, as opposed  
to transfer of mass into hotter phases, or to algorithmically determine  
the mass of the remnant of the original cloud. Bearing these complications   
in mind, we have simply totaled the mass  
of all cool gas (again $\log T < 4.08$) on the computational grid. For  
each model, comparison to the equivalent conductive model at the same  
time shows that  
for all cases there is more cool gas in the non-conductive models by a factor  
of typically $\la 15$\% in the low wind temperature models (T1 models),  
and $\sim 50$ to 100\% in the high wind temperature models (T5 models).   
Note that accounting for cloud fragments lost  
from the grid would increase these differences.  
In this crude sense mass loss (to the hot phase) is less in the non-conductive  
models. The higher total cool gas mass in the NC models is spread over multiple cloud fragments.  
 
For model T5LPNC we did identify the cloud core by hand  
at $t=1$ Myr and obtained a cold gas mass for this core 
of $47 M_{\odot}$. At this time the cold gas mass in conductive 
model T5LP is $\sim 100 M_{\odot}$.  
Thus in terms of maximizing the lifetime of the   
original cloud non-conductive models are not better than the conductive  
models.  
  
The simulations we have presented should be considered to as two extremes in  
which cloud lifetimes are minimized:  
clouds experiencing conduction-driven evaporation due to conduction at  
effectively saturated levels, verses  
efficient hydrodynamical fragmentation under conditions of   
no conduction at all. Our brief investigation of models 
with conduction at sub-Spitzer levels shows that in certain 
circumstances conduction is strong enough to  
suppress the hydrodynamical instabilities that fragment clouds, while 
also reducing cloud evaporation.  
As previously discussed larger clouds also have longer lifetimes. 
Thus it seems plausible that it will be possible to find conditions 
in which simulated cloud lifetimes are the $\ga 10$ Myr timescales 
we expected from observations, but further simulations 
will be required to test this hypothesis.

\subsection{Conduction-driven cloud implosion and star formation}  
\label{sec:discussion:implosion}  
  
The possible role of interstellar or inter-galactic shock waves in 
promoting star formation (or even galaxy formation) by crushing 
ambient gas clouds has been considered many times \citep[\eg 
see][]{woodward76,ostriker81,shchekinov89,begelman89,bicknell00,gorti02, 
fragile04,scannapieco04}.  Nevertheless, it has long been known that 
shock-crushing does have some undesirable effects, as the compression 
is not isotropic but tends to pancake the cloud \citep{woodward76}. 
  
The conduction-driven cloud compression seen in our simulations, and  
first noted by \citet{ferrara93}, should also be active in many of the   
contexts in which shock-driven cloud-crushing is considered, as long as  
hot gas is generated and thermal conduction is not completely suppressed.  
This conduction-driven compression is much more isotropic (due to  
the rapidity with which hot gas can wrap around a cloud) than   
shock crushing, and achieves high compression ratios.  
  
The clouds we have simulated are not natural candidates for star formation,  
given their low initial densities and masses. Nevertheless, conduction-driven  
implosion is a very interesting mechanism than deserves further investigation,  
and should be considered when thinking of extra-planar star formation in  
actively star-forming galaxies, and in galaxies over run by superwinds or  
jets.

\subsection{The origin and metallicity of the X-ray emitting gas}  
\label{sec:discussion:metal_abundances}  
  
What implications do these simulations have for understanding soft  
X-ray emission in superwinds? As discussed in Section \ref{sec:introduction},  
from observations it is clear that the majority of the   
soft X-ray emitting gas comes  
from a relatively small fraction of the total volume occupied by the wind.  
Current X-ray observational   
studies \citep[see][and references therein]{strickland04a} are  
consistent with the following scenarios: that the  
hot gas is ambient gas heated by the wind   
to X-ray-emitting temperatures, or alternatively  
is denser-than-average regions of initially hot merged SN and   
stellar wind ejecta, or is some mixture of the two. This is an  
important issue, for in hydrodynamical models of superwinds it is the  
merged SN ejecta that contains the majority of the   
energy of the wind \citep{ss2000}, and is the material most likely  
to escape into the IGM. If a significant fraction of the soft X-ray  
emission in superwinds comes from this material then future, higher-spectral  
resolution X-ray observatories, such as {\it Astro-E2} and   
{\it Constellation-X} (which can measure the velocity of the hot gas)  
will be able to directly constrain  
metal and energy ejection from starbursts into the IGM. If the hot gas is  
predominantly ambient ISM then these observatories will only probe a  
smaller fraction of the metal flow associated with superwinds.  
  
Observationally, the simplest diagnostic of the origin of the X-ray emitting  
gas would be to measure the absolute metal abundance. This has proved to  
be difficult, given a variety of systematic biases in the low to  
moderate resolution spectroscopy available with {\it ROSAT},   
{\it ASCA}, {\it Chandra}  
and {\it XMM-Newton}. These include  
the artificial blending of spectrally distinct regions   
\citep{dwh98,weaver00,dahlem00},  
calibration uncertainties  and over-simplistic  
spectral model fits to intrinsically complex spectra \citep{ss98}.  
The unphysically-low X-ray derived abundances found in the earliest Chandra  
observations of superwinds \citep[see \eg][]{wang01,strickland02,xia02}  
are almost certainly primarily due to small inaccuracies ($\la 2$\%)  
in the ACIS energy scale available  
at that time. Re-analysis of that data with more modern calibrations  
leads to significantly larger best fit abundances   
that are statistically inconsistent with the previous results,  
although it is still effectively impossible to distinguish between Solar or  
super-Solar element abundances with respect to hydrogen   
\citep[\eg][]{martin02,strickland04a}. In the near future observations 
with the much higher spectral resolution provided by the 
{\it Astro-E2} calorimeter will not suffer from these problems, 
and so should provide strong constraints on the absolute elemental 
abundances in the nearest bright superwinds.

We discussed the origin of the soft X-ray emission in our models  
of a wind/cloud interaction in Section \ref{sec:results:emission}.   
Here we restate those  
claims and develop our analysis further, by also considering cases where 
 $Z_{\rm w} > Z_{\rm c}$.

Soft X-ray emission in the cloud models without thermal   
conduction (the NC models) generally arises in wind   
material within the bow shock (see Fig.~\ref{fig:ral55emi}),  
with a model-dependent contribution from the cloud surfaces.  
In the models with conduction the soft X-ray luminosities are  
typically larger (by a factor between 2 and 30)   
than in the non-conductive models (Table~\ref{tab:percentual_luminosity}),  
and this increased luminosity is due to the increase in 
the hot gas density in the region between the cloud surface and the bow shock  
due to evaporated cloud material (see Fig.~\ref{fig:rar55emi}).  
  
Note that the O{\sc vi} luminosities of both classes of model  
are very similar.  
In these models the gas emitting O{\sc vi} typically has a density of $n  
\sim 5 \times 10^{-2}$ cm$^{-3}$, and is always material originally from the  
cloud in both conductive and non-conductive models.  
As such it will have the metallicity  of the unenriched  
cloud, \ie low abundances if the cloud is a halo or   
high velocity clouds over-run by the wind, or the metallicity of  
the WIM of the host galaxy if it is being advected out of the disk within  
the wind.

 
\begin{figure*}   
\begin{center}   
\psfig{figure=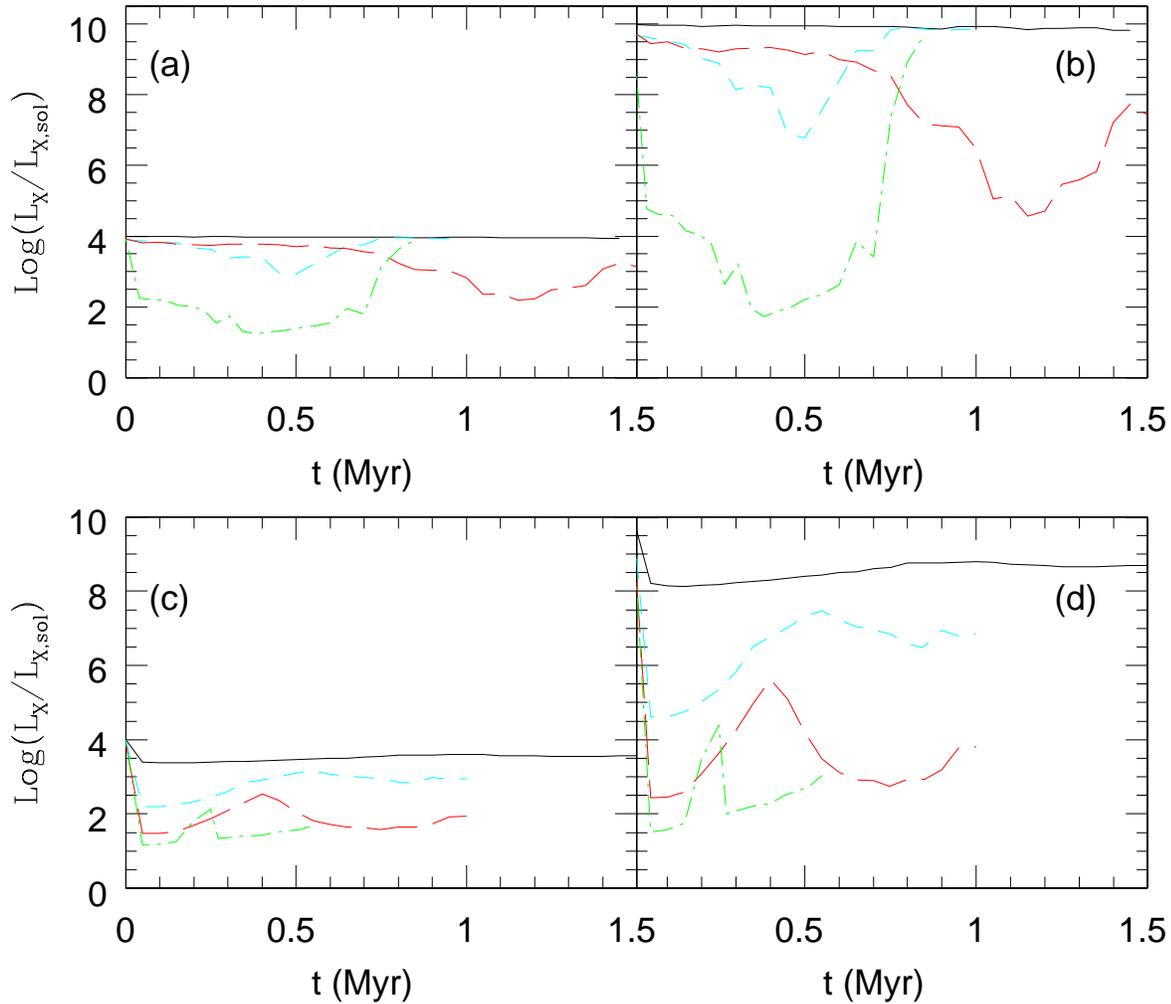}
\end{center}   
\caption{X-ray luminosities for 
high $Z_{\rm w}$ normalized to $L_{\rm X}$ obtained with $Z_{\rm w} = 1
Z_{\odot}$. Black solid lines: T1LP(NC); blue short dashed: T1HP(NC); red 
long dashed: T5LP(NC); green dot dashed: T5HP(NC). In all cases $Z_{\rm c} = 1
Z_{\odot}$.  (a) Non-conductive models, assuming $Z_{\rm w} = 4
Z_{\odot}$.  (b) Non-conductive models, assuming $Z_{\rm w} = 10
Z_{\odot}$.  (c) Conductive models, assuming $Z_{\rm w} = 4
Z_{\odot}$.  (d) Conductive models, assuming $Z_{\rm w} = 10
Z_{\odot}$.}
\label{fig:highz_lx}  
\end{figure*}  
 
The results presented in Section \ref{sec:results:emission}  
were made under the assumption 
that the metal abundance of the wind and cloud material were both 
Solar. As discussed in Section \ref{sec:numerical_method}, the metal 
abundance of the wind material can be considerably higher than 
Solar, depending on the degree of wind mass-loading. In a situation 
where $Z_{\rm w} > Z_{\odot}$ wind material will make up a larger 
fraction of the X-ray emission, as the 
soft X-ray emissivity scales roughly in proportion to the metal abundance 
for gas in the temperature range $T \sim 10^{6}$ -- $10^{7}$ K. 
Note that there must be little or no mass loading of the wind 
for its metal abundance to be high. Consequently the wind velocity and 
temperature will be higher and wind density lower  
than in a heavily mass-loaded case. 
Thus our T1 models could be considered as representing heavily mass loaded 
winds, and the T5 models as non-mass loaded or weakly mass loaded winds. 
 
To illustrate the influence of the wind metal abundance on 
the emission from the wind/cloud interaction, we recomputed the  
X-ray luminosity as function of time in all models assuming that 
the X-ray emissisivity of all wind material was either 4 or 10 times 
higher than for a Solar abundance plasma. In Fig.~\ref{fig:highz_lx} 
we show the ratio of this luminosity $L_{\rm X}(Z)$ to the 
X-ray luminosities calculated assuming $Z_{\rm w}= Z_{\odot}$ 
(which were shown in Fig.~\ref{fig:emi}) as a function of time.  
Changing the abundance of the wind 
material will have negligible effect on the O{\sc vi} emission, 
which is always dominated by cloud material. 
 
The large increase in X-ray luminosity seen in the T5LPNC models 
for high $Z_{\rm w}$ illustrates the dominance of wind material 
in this model (long dashed lines in Fig.~\ref{fig:highz_lx}a and b). 
Averaged over the entire length of the simulations wind fluid material 
accounts for 35\% of the soft X-ray emission when  
$Z_{\rm w} = 1 Z_{\odot}$. This fraction increases to 88\% (95\%) if  
$Z_{\rm w} = 4 Z_{\odot}$ ($Z_{\rm w} = 10 Z_{\odot}$). 
In contrast, with conduction active the wind material only accounts 
for on average 25\% of the X-ray emission in model T5LP when 
$Z_{\rm w} = 1 Z_{\odot}$. Although the increase in X-ray 
luminosity for this model at high $Z_{\rm w}$ (see  
Fig.~\ref{fig:highz_lx}c and d) is significantly less than in model 
T5LPNC, wind material does dominate the emission (58\% of the emission 
for $Z_{\rm w} = 4 Z_{\odot}$, 77\% of the emission for  
$Z_{\rm w} = 10 Z_{\odot}$). 
 
All the T1 models are on average dominated by wind material, for all values 
of $Z_{\rm w}$, so the X-ray luminosity does increase strongly 
with increased $Z_{\rm w}$. 
 
The T5HP models are another interesting case. Despite the very different 
dynamical state of the T5HPNC and T5HP models, the average fraction 
of the X-ray emission from wind material in both models is relatively 
similar (although always slightly larger in the non-conductive NC models, 
representing the lack of any contribution due to strong evaporative loading of 
the bow shock region that is seen in the conductive model). For 
$Z_{\rm w} = 1 Z_{\odot}$ wind material only produces 12\% of the emission  
in the T5HP model, and 14\% of the emission in the T5HPNC. As argued  
previously a wind with conditions similar to those in the T5HP(NC) models 
can not be heavily mass-loaded, and hence will have a high metal abundance. 
Cloud material dominates in both of these models for  
$Z_{\rm w} = 4 Z_{\odot}$, but 
for $Z_{\rm w} = 10 Z_{\odot}$ wind material 
provides 57\% of the emission   
in the T5HP model and 66\% of the emission in model T5HPNC.

Our discussion using broad energy band weighted luminosities and 
${\cal R}$ ratios is   
intended primarily as an aid to understanding, but some of the  
subtleties and second order effects are also worth considering.  
Not only does the relative degree of enrichment   
of the wind (or $Z_{\rm SN}/Z_{\rm ISM}$)  
vary from element to element,   
but the degree of contribution of wind material (${\cal R}$) to the 
X-ray emission  
is also element dependent, given the different temperature  
of the conductive interface compared to the bow shock.   
The emissivity of  
different elements and their ions are strongly temperature dependent.  
In these simulations the conductive interface has a lower temperature  
than the wind bow shock, and thus  
emission from   
lower-temperature ions (\eg O{\sc vii} or O{\sc viii}) will be more  
heavily influenced by cloud material than higher temperature  
ions (\eg Mg{\sc xi}, Mg{\sc xii} or Si{\sc xiii}). In some cases,  
particularly when conductive-interface and wind contributions are  
on average comparable, this effect would lead to peculiar   
X-ray-derived abundance patterns, with lower ionization state ions   
appearing to have lower abundances than high-ionization ions.   
It is conceivable that the peculiarly-low X-ray-derived ratio   
of Oxygen to other $\alpha$-element  
abundances found in M82 by \citet{origlia04} might be caused by such  
an effect, although there are a few other observational biases   
that could also cause these odd results.

The point that should be taken away is   
the gas probed by soft X-ray observations of superwinds  
is not necessary always un-enriched ambient gas even when thermal  
conduction is active at the high Spitzer-conductivity levels. For some  
sets of conditions (and counter to na\"{\i}ve expectation), in particular in
powerful high velocity and high pressure winds, the ambient gas being  
conductively evaporated off the cloud does dominate the observable soft   
X-ray emission. In less powerful winds conductive evaporation is reduced,  
and in these case and in the case of no conduction, the soft X-ray  
emission arises in wind material within the bow shock around each cloud.  
The metal abundance of the wind material  
depends strongly on the degree of mass-loading of the volume-filling  
wind-fluid, which is a separate topic deserving some additional investigation. 
However, this implies that if we can measure the metal abundance of the 
soft X-ray emitting gas it can be representative of the more tenuous 
volume-filling wind, and thus can be used to constrain the degree of 
large-scale wind mass loading.

\subsection{Soft X-ray and O{\sc vi} emission and absorption}  
\label{sec:discussion:luminosities}  
  
\subsubsection{O{\sc vi} absorption}  
  
The mean O{\sc vi} column densities through a   
simulated single wind/cloud interaction (Table~\ref{tab:column_density})  
are typically 1.5 dex lower than the value of  
$N_{\rm O VI}$ seen in starbursts observed with {\it FUSE}, which  
range from $14.3 < \log N_{\rm O VI} < 15.3$ \citep[see][]{heckman02}.  
Increasing the cloud radius, or the metal abundance of the wind, is not  
likely to increase $N_{\rm O VI}$, for the reasons discussed in  
Section \ref{sec:results:cloud_radius} and   
Section \ref{sec:discussion:metal_abundances}.

There is relatively  
little difference in mean $N_{\rm O VI}$ between the conductive  
and non-conductive models. While it is hard to assess to what degree numerical  
diffusion is artificially increasing the amount of O{\sc vi}   
absorbing material in the non-conductive models, it is certainly  
true that $N_{\rm O VI}$ is not currently a strong discriminant between  
conductive and non-conductive models for wind/cloud interactions.  
  
The hotter and/or faster wind models produce larger mean $N_{\rm O VI}$  
per cloud. A high mean $N_{\rm O VI}$ per cloud is advantageous,  
in that it reduces the number of clouds needed to match observations.  
Unfortunately, cloud stripping or evaporation is more efficient in  
such models, so either more clouds would be required initially, or a  
continuous source of clouds would be required  
(\eg via Kelvin-Helmholtz instabilities along the walls of the cavity).  
  
There would need to be $\sim 30$ clouds per line of sight if  
clouds alone are responsible for the O{\sc vi} absorption in superwinds.  
In practice the walls of the outflow will be another significant  
source of O{\sc vi} absorption.  
That there should be multiple clouds along any line of sight through  
a superwind is by no means surprising, given that the velocity  
width of observed optical and UV absorption lines is much greater than  
could be expected from any single structure. Nevertheless, a   
model requiring very large numbers of clouds along any line of sight  
is neither parsimonious nor elegant. In summary, the simulated   
O{\sc vi} column densities alone do not provide strong constraints on  
allowable wind models when compared to the existing observations.  

As shown in \citet{heckman02} superwinds follow a distinct O{\sc vi}  
column density vs line width ($b$) relationship.  
We would expect that the column density and 
line width of a simulated 
O{\sc vi} profile from a single cloud should fall on the trends 
shown in Fig.~1 of that paper if a superposition of clouds is to 
create the net profile observed. 
Of the simulated O{\sc vi} profiles we investigated only 
half of them fell on the low $N_{O VI}$, low $b$ extrapolation 
of the \citet{heckman02} trend. The other simulated profiles 
had low $N_{O VI}$ but high $b$ values of $\sim 200$ km/s. 
This is puzzling, but it is too early to assess if this is a 
significant discrepancy.  
Our simulations only cover the first 1 Myr of a cloud/wind interaction, 
while real superwinds should contains clouds with a broad range of 
ages.

\subsubsection{O{\sc vi} and soft X-ray emission}  
  
In Table~\ref{tab:obs_flux_ratios} we summarize the published values of  
O{\sc vi} and soft X-ray fluxes in three starburst-driven  
superwinds (M82 and NGC 3079 host strong superwinds, while 
NGC 4631 is a weaker starburst) and the marginal O{\sc vi} detection in the  
halo of a normal spiral galaxy believed to have a galactic fountain  
(NGC 891), adapted from the work of \citet{hoopes03} and \citet{otte03}.   
We choose to concentrate  
our analysis on the ratio of O{\sc vi} to soft X-ray emission,  
and the ratio of the emitted FUV and X-ray power to the wind mechanical  
power, given that we can only simulate a single cloud. The $30\arcsec$-wide  
aperture of the FUSE observations  
samples a small fraction of the total region covered by soft X-ray  
emission in these objects, but this is considerably larger than volume  
we can simulate with high resolution.  
  
Note even with the limited observational data currently available it
is clear that there is not a single universal $\lo/\lx$ ratio in
superwinds: the upper limits on the ratio in M82 and N3079 are
significantly lower than the values based on a statistically strong
detection of O{\sc vi} in the NGC 4631 wind.
  
The value of the wind mechanical luminosity intercepted by the cloud,
used in Table~\ref{tab:percentual_luminosity}, is calculated over a
radius of 10 pc, which is roughly the mean radius the compressed or
crushed clouds assume. Note that the X-ray and O{\sc vi} luminosities
are calculated using a larger radius of 50 pc (in order to encompass
the majority of the bow shock around the cloud). This mismatch in
radii is reasonable, as the X-ray emission region is always larger
than the cloud itself.  If we were to use a 50 pc radius for both
$L_{\rm mec}$ and the emitted radiation then the mean $\lo/L_{\rm
mec}$ and $\lx/L_{\rm mec}$ ratios would be lower by a factor 25.
  
By using a radius equivalent to the effective cloud radius in calculating  
$L_{\rm mec}$, and hence the $\lo/L_{\rm mec}$, $\lx/L_{\rm mec}$  
and $\L_{\rm tot}/L_{\rm mec}$ ratios, we are effectively assuming the  
cloud area covering factor (as seen by the wind) is of order unity.   
If the cloud covering factor is less than unity, then the mean  
radiative to mechanical energy ratios should be reduced proportionally.  
However, it should be noted that   
FUV spectra of starbursts taken with {\it FUSE} indicate that  
the area covering fraction of neutral atomic clouds (\eg seen in   
C{\sc ii} $\lambda 1036$ absorption) is  
indeed of order unity  
\citep{heckman01b}, in contrast to the lower  
 area covering fraction of (presumably denser)  
molecular gas clouds \citep{hoopes04a}.

Table~\ref{tab:percentual_luminosity} shows the mean fraction of  
wind mechanical energy converted in X-ray or FUV radiation \emph{by a  
single cloud} of initial radius 15 pc, along with the mean  
O{\sc vi} to soft X-ray flux ratio. As O{\sc vi} emission accounts  
for $\sim30$\% of the total emissivity of gas at $T\approx 3\times10^{5}$  
K, we can combine the O{\sc vi} luminosities with the  
soft X-ray luminosities to gain an estimate the of total   
radiative energy loss from the coronal and hot phases.

We find that a  
low fraction of the wind mechanical energy intercepted by a single  
cloud is lost to radiation. 
The total fraction of wind energy radiated per cloud found in the  
models assuming $Z_{\rm w} = 1 Z_{\odot}$  
lies in the range $\sim 0.25$ -- $1$\%, and is relatively  
similar between the conductive and non-conductive models.  
For higher values of $Z_{\rm w}$ the ratio $\lx/L_{\rm mec}$  
per cloud is still $\la 1$\%, except in 
the T1LP and T1LPNC models where it reaches 2 -- $3$\%. 
As mentioned previously a somehwat larger fraction of wind energy 
ends up as cloud kinetic energy. 
 
The simulations show that a superwind can  
flow around a cloud and thus affect multiple clouds along any particular  
path without any substantial radiative loss of wind energy, except in the case 
of the T1LP(NC) models where radiative losses may violate observational 
limits.  
Summed over all clouds the $L_{\rm tot}/L_{\rm mec}$  
ratio should not significantly exceed  
estimates of  the X-ray and FUV luminosity to wind mechanical power  
in real superwinds. Using {\it FUSE} observations of the dwarf starburst  
NGC 1705, \citet{heckman01} limit  
to the coronal cooling rate to be $\sim 5/\epsilon$\% of the mechanical  
luminosity based on the observed O{\sc vi} absorption line strength.  
Soft X-ray observations of superwinds give a ratio of $\sim 1/\epsilon$   
-- $3/\epsilon$\% for the X-ray to mechanical power ratio   
\citep{strickland_brazil}. The major uncertainty in these   
observationally-derived ratios is  
the unknown efficiency of supernovae energy thermalization $\epsilon$.  
  
The lower pressure wind models (LP models) produce more emission per unit  
mechanical power than the high pressure models, although the   
magnitude of $\lo$ or $\lx$ is larger per cloud in the high  
pressure models. The lower temperature wind models (all T1 models) also  
produce more emission per unit $L_{\rm mec}$ than the higher  
temperature wind models, but this is  
mainly because of the high wind density in these models.

In accord with the observational evidence, coronal phase cooling (\ie
emission predominantly in the FUV) dominates over X-ray cooling from
hotter gas by a factor of at least a few. The strongest difference
between the models consists in the X-ray cooling efficiencies. With
the one exception of model T1LPNC, the $\lx/\L_{\rm mec}$ ratios are
an order of magnitude lower in non-conductive models compared to the
models with conduction.

Another significant difference between the conductive and non-conductive  
clouds models is in the O{\sc vi} to X-ray luminosity ratio. 
In our opinion this ratio places the 
most interesting constraints on whether wind/cloud interaction models 
can reproduce observed superwind properties. 
 
For $Z_{\rm w} = Z_{\odot}$ the majority  
of the non-conductive models have $\lo/\lx$  
ratios that are orders of magnitude larger than the largest  
values observed to date (compare   
Table~\ref{tab:percentual_luminosity} to Table~\ref{tab:obs_flux_ratios}).  
The low pressure non-conductive model T1LPNC has $\lo/\lx = 1.07$, but  
this value is still much larger than the upper limits on the   
$\lo/\lx$ ratio for M82 and NGC 3079.

The conductive models span a much smaller range in $\lo/\lx$, from 0.2  
to 2.7. Again, the low temperature low ram pressure models have  
the lowest $\lo/\lx$ ratios, closer to the observed range, if still  
only matching the values observed in NGC 4631. 
  
The fundamental problem with the  
non-conductive models are the low X-ray luminosities.  
The O{\sc vi} luminosities are comparable in both conductive  
and non-conductive models, so it is the higher soft X-ray luminosities   
generated by cloud material in the conductive interface that  
leads to roughly acceptable  $\lo/\lx$ ratios. It  
may be the case that  
the typical $\lo/\lx$ ratio in superwinds is $\la 0.1$,  
in which all of our wind/cloud interaction models have trouble producing   
a realistic $\lo/\lx$ ratio if $Z_{\rm w} \sim Z_{\rm c}$.

Larger clouds are more effective at intercepting wind energy and  
converting it into radiation. The mechanical power intercepted scales  
as $R_{\rm c}^{2}$, but the X-ray emission scales as roughly   
$R_{\rm c}^{3}$ over the small range of radii we  
have investigated (see Section \ref{sec:results:cloud_radius}).  
The O{\sc vi} luminosity does not grow with increased  
radius as rapidly as the X-ray luminosity, so the $\lo/\lx$ ratio  
per cloud will drop as for clouds of larger radius. This effect  
is probably not significant enough to reconcile 
the low $Z_{\rm w}$ non-conductive models with observed $\lo/\lx$  
ratios, but should be considered for the conductive models.  
In the case of models with thermal conduction, matching the 
 $\lo/\lx$ ratio in M82 and NGC 3079 is possible if the average 
cloud has an initial radius only a few times larger than $15$ pc.  
Larger cloud radii also 
have the added advantage of longer cloud lifetimes, as previously 
discussed. 
 
Even values of $Z_{\rm w} = 10 Z_{\odot}$ are not sufficient to produce 
acceptably low $\lo/\lx$ for the non-conductive models  
(see Table~\ref{tab:percentual_luminosity}). However, for the 
conductive models a  high $Z_{\rm w}$ 
value does produce $\lo/\lx$ values similar to the 
$\lo/\lx \la 0.1$ upper limit in M82 and NGC 3079. 
Thus even in the case of $Z_{\rm w} \gg Z_{\rm c}$ the  
O{\sc vi} to X-ray ratio appears to provide a strong discriminant 
between non-conductive and conductive cloud models for superwinds, 
with conductive models producing values in the range of those observed. 
 
Furthermore, it appears that we need not require conduction at 
full Spitzer levels to attain acceptably low  
$\lo/\lx$ ratios.  As shown in Table~\ref{tab:column_density_reducted} 
a reduction in the value of the coefficient of thermal 
conductivity by a factor $f=5$ barely altered the 
time-averaged $\lo/\lx$ ratio. Even for a reduction of  
$f=25$ the increase in $\lo/\lx$ is surprisingly small. Conduction 
is still dynamically important in these models in stabilizing the 
cloud against fragmentation, so it is possible that with 
weak to moderate conduction we can produce longer-lived clouds with 
suitably low $\lo/\lx$ ratios. Further simulations tracking clouds 
over significantly longer timescale will be necessary to test this hypothesis. 
 
For the set of parameters investigated in this paper 
the models T1HP and T5LP have the most acceptable 
X-ray and O{\sc vi} emission properties, in that the  
$L_{\rm tot}/L_{\rm mec}$ ratio per cloud is not too large 
(even in cases where $Z_{\rm w} \sim 10 Z_{\odot}$) and 
the $\lo/\lx$ ratio is also low. 
  
\section{Summary}  
\label{sec:conclusions}  
  
We performed a series of 2-dimensional  
hydrodynamical simulations with a resolution $0.1$ pc of a   
super-or-Hyper-sonic  
starburst-driven superwind interacting with an initially  
stationary cool dense cloud of radius $R_{\rm c}=15$ pc.   
Such interactions are one of a class of wind/cloud  
interactions known to occur within superwinds.   
Wind/cloud interactions may play a significant role in shaping   
their observable properties, in particular in the FUV and X-ray energy bands,  
but have not been explored numerically in any detail.  
More general analytical models   
often applied to superwinds, such as the \citet{weaver77}  
wind-blown bubble model, have trouble explaining observed O{\sc vi}  
absorption line column densities and O{\sc vi} to X-ray emission flux  
ratios.  
  
We considered a range of models,  
with low and high wind ram pressure (both  
pressures are much higher than conventional ISM pressures),   
low and high wind to cloud density ratios,  
and the presence or absence of thermal conduction. The resolution  
of the simulations is high enough that the region of the conductive interface  
in which the strong coolant O{\sc vi} is produced is numerically resolved.  
A numerical tracer field was used to distinguish between gas originally  
part of the cloud and gas from the wind.  
We followed the  
evolution of the wind/cloud interaction for times up to 1.5 Myr, at  
which point the cloud has been carried off the numerical  
grid by the wind. In addition to considering the dynamics of  
the wind/cloud interactions, we have also modeled their soft  
X-ray emission and FUV O{\sc vi} emission and absorption properties  
and compared them to existing observational data.  
  
There are significant dynamical differences between wind/cloud interactions  
in the presence of  
 thermal conduction compared to those without conduction. In the  
absence of conduction the clouds are fragmented in a few dynamical  
(cloud-crushing) time-scales $\tau_{\rm cc} = \chi^{1/2} \rc / \vw$ (where  
$\chi=\rho_{\rm c}/\rho_{\rm w}$),   
as expected from earlier of clouds  
over-run by SN blast waves \citep{klein94}. Relatively little  
cool cloud mass is converted into hot gas by shocks or mixing,   
but the cloud rapidly  
ceases to exist as a recognizable coherent object due to the   
hydrodynamical instability-driven fragmentation.  
  
In contrast, thermal conduction acts to stabilize the cloud by
inhibiting the growth of Kelvin-Helmholtz and Rayleigh-Taylor
instabilities (as had previously been found in the case sub-sonic
motion of Molecular clouds in a hot medium by
\citealt{vieser00}). Furthermore, the conductive heat flux into the
cloud's surface generates a shock wave that compresses the cloud and
modifies its internal density structure significantly and reduces its
conductive mass loss rate (as previously discovered for stationary
clouds by
\citealt{ferrara93}). These dynamical effects occur even in models 
in which the coefficient of thermal conductivity has been reduced by
factors of up to 25 from the Spitzer value.  If thermal conduction is
active, then clouds embedded within a superwind will exist as a single
coherent structure for longer time scales than they otherwise would,
\emph{but} at the cost of losing mass to evaporative heating. The
strength of these conductive effects is proportional to the
dimensionless parameter $\sigma_{0}$ (Equation~\ref{equ:sigma0}). This
parameter is high for low density winds, small clouds, and high wind
temperature, especially in the case of supersonic winds where the
temperature of material in the bow shock around the cloud is greater
than the wind temperature.
  
In general, the higher the wind ram pressure or velocity is, the higher the  
mass loss rate of a cloud will be, due to either hydrodynamical  
fragmentation or conductive evaporation.   
  
With or without the presence of thermal conduction, O{\sc vi} emitting
and absorbing material primarily arises in cloud material at the
periphery of the cloud and any cloud fragments. In contrast, the
origin of the soft X-ray emitting material depends on the strength of
conductive effects and the relative metal abundances of the wind and
cloud material.  If conductive effects are weak or absent, and/or the
metal abundance of the wind $Z_{\rm w}$ is highly enriched compared to
the cloud $Z_{\rm c}$, then the observed soft X-ray emission comes
from wind material in the bow shock around the cloud. Thus even if the
undisturbed superwind is relatively tenuous and thus a weak X-ray
emitter, when compressed around an obstacle such as a cloud its
emission is enhanced and hence the wind abundance pattern could be
probed by soft X-ray observations.  For strong conduction, or in the
case of weaker conduction with $Z_{\rm w} \sim Z_{\rm c}$ (\eg a
heavily mass-loaded superwind) the X-ray emission again arises within
the bow shock region, but conductively evaporated cloud material mixed
into this region dominates the emissivity.  In the highest wind ram
pressure and wind temperature models with conduction or without
conduction it is still cloud material that dominates the X-ray
emission, even if the metal abundance of the wind is 10 times greater
than that of the cloud.
  
Thus whether soft X-ray observations probe wind or cloud material  
in a wind/cloud interaction  
depends on several factors: primarily the strength of conduction  
(conduction favouring cloud material) and the relative abundances  
of wind with respect to the cloud (if the wind is strongly   
enriched then its influence increases). Different material can  
in principle been seen at different X-ray energies, with low  
ionization species such as Oxygen being more strongly weighted  
toward probing the cooler conductive interface than high ionization  
species such as Mg or Si (which would tend to probe the hotter regions  
of the bow shock). This complexity may contribute to some of the  
uncertainties and peculiarities found in current  
X-ray-derived metal abundances in starbursts  
\citep[\eg][]{origlia04}.  
  
The O{\sc vi} absorption line column density produced per cloud  
is relatively similar between conductive and non-conductive models,  
ranging from $12.9 \la \log N_{\rm O VI} \la 13.5$.  
In all cases the mean $N_{\rm O VI}$ value found is substantially  
less than the observed total O{\sc vi} column densities in starbursts  
with winds, so that in the wind/cloud interaction model requires  
many clouds per line of sight (typically 10 -- 30 clouds).  
Models with stronger hydrodynamical stripping or conductive evaporation  
produce higher mean $N_{\rm O VI}$, but their shorter cloud lifetimes  
are disadvantageous for accumulating larger numbers of clouds,  
and maintaining cool cloud gas populations in winds for the  
$\sim 10$ Myr time periods that appear to be implied by observations.  
  
In terms of emission we considered two primary observational  
constraints. The strongest observation constrain is the ratio between  
O{\sc vi} and soft X-ray emission $\lo/\lx$, which has the value   
$\sim 1$ in NGC 4631 \citep[this work, see also][]{otte03},  
 and is definitely much lower ($\la 0.1$)  
in the stronger M82 and NGC 3079 superwinds  
\citep[ Hoopes et al, in preparation]{hoopes03}.   
The second is that observations  
place limits on the fraction of wind mechanical power that  
is radiated in the FUV and X-ray bands, which are relatively low  
(\eg $\la 5/\epsilon$\%) but made somewhat uncertain by our  
lack of robust knowledge of the starburst SN thermalization  
efficiency $\epsilon$. We found that except for the low ram pressure low wind 
temperature T1LP(NC) models, $\la 1$\% of the wind mechanical 
energy is lost to radiation per cloud, even for wind metal abundances 
as high as $10 Z_{\odot}$. Typically between 2 -- 6\% of the wind 
mechanical energy ends up as kinetic energy of the cold cloud within 
the first 0.5 Myr.  
  
If we consider both possible cases, of both Solar and super-Solar  
abundance winds, we can draw the following conclusions.  
Qualitatively, the high wind-ram pressure models or the 
low ram pressure high wind temperature model are somewhat   
better than the low pressure, low wind temperature, models  
at matching the available constraints on X-ray emission  
and O{\sc vi} emission and absorption.  
If $Z_{\rm w} \sim Z_{\rm c}$ then the non-conductive  
models are ruled out by their very high $\lo/\lx$ ratios (due to the  
relative lack of X-ray emission), while  
conductive models can match the value of $\sim 1$ observed in  
NGC 4631. Even if $Z_{\rm w} = 10 Z_{\odot}$ then  
the $\lo/\lx$ ratio in most of the non-conductive models remains 
higher than the value observed in NGC 4631.  
For high $Z_{\rm w}$ the conductive models produce $\lo/\lx$ 
as low as $\sim 0.1$, similar to the upper limits on the ratio 
in M82 and NGC 3079. 
 
Thus some, but not all, of our current wind/cloud interaction models can 
match the range of $\lo/\lx$ values observed in the starbursts 
NGC 4631, M82 and NGC 3079, if conduction is active, and/or with  
high wind metallicity (\ie little wind mass loading). 
The low wind ram pressure, low wind temperature models (T1LP/NC) 
radiate too large a fraction of the wind's mechanical energy  
per cloud to allow large numbers of clouds along any line of sight, 
which may rule them out as a plausible model. 
A smaller fraction of the superwind's   
energy is radiated per cloud in the high pressure models,   
but the total luminosity per cloud  
is slightly higher. This is advantageous, as  fewer clouds  
are required instantaneously, but the ultimate issue of cloud  
lifetime and replacement still remains. Larger clouds 
or lower thermal conductivity are plausible ways in which 
cloud lifetimes can be extended to the $\sim 10$ Myr time scales 
required by observations. 
 
Wind/cloud interactions in superwinds have been discussed
qualitatively for a long time, stretching at least as far back as the
seminal paper of \citet{chevclegg}. The conditions that appear to
favor long cloud life times are not necessarily those that appear to
produce the best match to FUV and X-ray emission and absorption
properties, but it does appear that these competing requirements can
be balanced.  More work is required to follow clouds for longer
periods, and to better investigate cloud acceleration and thermal
conduction at sub-Spitzer levels.

That there is some level of diversity in the physical  
conditions within different starburst-driven superwinds 
and thus in their emitted radiation is not totally unexpected.   
In principle this can be advantageous if we are  
to disentangle the multiple fluids variables that shape a superwind  
and understand their physics. Observing more superwinds in  
detail, especially with {\it FUSE}, will be scientifically profitable.  
  
Nevertheless, we find it  very encouraging that some of these  
wind/cloud models can quantitatively   
match some of the observed properties of winds.  
Furthermore, these simulations have been very enlightening with  
regard to the both the role conduction may play in winds (a topic that  
has seen very little direct work despite its obvious potential  
importance), and to what controls which material (cloud or wind) is probed by  
X-ray observations. This latter issue is apposite given   
the immanent launch of the very high spectral resolution  
X-ray micro-calorimeter on {\it Astro-E2}. 
With direct observational probes of the elemental composition 
and possibly the kinematics of the soft X-ray-emitting gas in 
superwinds it will be possible to constrain the physical 
processes that shape emission from these outflows. 
  
\section*{Acknowledgments} 
We are grateful to the referee for his/her helpful suggestions which
improved the presentation of the paper. We acknowledge financial
support from National Institute for Astrophysics (INAF). The
simulations were run at the CINECA Supercomputing Centre with CPU time
assigned thanks to INAF-CINECA grant. One of us (A.M.) thanks the
Johns Hopkins University for hospitality during work on this paper.
  
\bibliographystyle{mn2e}  
\bibliography{MF480_refs}  
  
\label{lastpage}   
\end{document}